\documentclass[11pt]{article}
\makeatletter
\@addtoreset{equation}{section}

\DeclareSymbolFont{boldgr}{OML}{cmm}{b}{it}
\DeclareSymbolFontAlphabet{\bi}{boldgr}
\DeclareSymbolFont{boldcl}{OMS}{cmsy}{b}{n}
\DeclareSymbolFontAlphabet{\bcal}{boldcl}
\DeclareMathAlphabet{\Bbb}{U}{msb}{m}{n}
\DeclareMathAlphabet{\euf}{U}{euf}{b}{n}
\DeclareMathSymbol{\bDelt}{0}{boldgr}{"01}
\DeclareMathSymbol{\bLamb}{0}{boldgr}{"03}
\def\eqlabel[#1]{\label{#1}}
\def\beq{\arraycolsep .1em\begin{eqnarray}\@ifnextchar[{\eqlabel}{}}
\def\eeq{\end{eqnarray}}
\def\zl{\nonumber\\}
\def\txt#1{\quad\hbox{#1}\quad}

\def\eref#1{(\ref{#1})}

\def\i{\mathrm{i}}
\def\d{\mathrm{d}}
\def\^{{}^}
\def\_{{}\lowind3_}
\def\lowind#1{\@ifnextchar_{\makebox[-.#1ex]{ }}{}}
\def\cc{^*\lowind9}
\def\dgg{^\dagger\lowind6}
\def\intdt{\int\!\d t\,\,\,}
\def\intdtau{\int\!\d\tau\,\,\,}
\def\intdsig{\int\!\d\sigma\,\,\,}
\def\intd#1#2{\int\!\d^{#1}#2\,\,\,\,}
\def\intdk{\int\!\frac{\d^3k}{(2\pi)^3}\,\,}
\def\intdkkk{\int\!\frac{\d^3k}{2 (2\pi)^3}\,\,}
\def\intdkk{\int\!\frac{\d^3k}{(2\pi)^3|k|}\,\,}
\def\sumk{\frac 1{2\pi}\sum_k\,\,}
\def\suml{\frac 1{2\pi}\sum_l\,\,}
\def\sumll{\frac 1{2\pi}\sum_{l>0}\,\,}
\def\sumlll{\frac 1{\pi}\sum_{l>0}\,\,}
\def\diag{\mathrm{diag}}
\def\sign{\mathrm{sign}}
\def\conf{{\cal Q}}
\def\phas{{\cal P}}
\def\phys{{\cal S}}
\def\Ext{\mathrm{Ext}}
\def\con{{\psi}}
\def\fcl{{\phi}}
\def\scl{{\chi}}
\def\pmat{{\bDelt}}
\def\imat{{\bLamb}}
\def\q{{\bi q}}
\def\v{{\bi v}}
\def\u{{\bi u}}
\def\p{{\bi p}}
\def\x{{\bi x}}

\def\qq{q'{}\lowind7}

\def\ba{m}
\def\bb{n}
\def\bc{p}
\def\op#1{\widehat #1}
\def\hilb{{\cal H}}
\def\stsp{{\cal V}}

\def\pot{V}
\def\obs{O}
\def\dt#1{\mathbf{\dot{\mathnormal{#1}}}}
\def\follows{\quad\Rightarrow\quad}
\def\equiv{\quad\Leftrightarrow\quad}
\def\ft#1#2{{\textstyle{{#1}\over{#2}}}}
\def\intd#1#2{\int\!\d^{#1}\!#2 \,}

\def\weak{\approx}
\def\comm#1#2{{\big[#1,#2\big]}}
\def\pois#1#2{{\big\{#1,#2\big\}}}
\def\dirac#1#2{{\pois{#1}{#2}{}_*}}
\def\del{\partial\lowind1}
\def\deldel#1/#2/{\frac{\del #1}{\del #2}\,}
\def\dd#1/#2/{\frac{\d #1}{\d #2}\,}
\def\deltadelta#1/#2/{\frac{\delta #1}{\delta #2}\,}
\def\eps{\varepsilon}
\def\RR{\Bbb{R}}
\def\act{I}
\def\acth{\tilde I}
\def\lag{L}
\def\ham{H}
\def\phm{H_0}
\def\bra#1{\big\langle#1\,\big|\,}
\def\br#1{\langle#1|}
\def\ket#1{\,\big|\,#1\big\rangle}
\def\kt#1{|#1\rangle}
\def\braket#1#2{\big\langle #1 \,\big|\, #2 \big\rangle}
\def\brkt#1#2{\langle #1|#2\rangle}
\def\expect#1{\big\langle\, #1 \,\big\rangle}
\makeatother

\begin{document}

\title{Dirac's Canonical Quantization Programme}
\author{Hans-J\"urgen Matschull\\[3mm]
       Mathematics Department, Kings College London,\\
       London WC2R 2LS, England}
\date{May 1996}

\maketitle

\begin{abstract}
This is a collection of lectures given at the University of Heidelberg. 
They are about Dirac's general method to construct a quantum theory out of 
a classical theory, which has to be defined in terms of a Lagrangian. The 
classical Hamiltonian formalism is reviewed, with emphasis on the relation 
between constraints and gauge symmetries, and quantization is carried out 
without any kind of gauge fixing. The method is applied to three examples: 
the free electro-magnetic field, the relativistic point particle, and the 
very first steps of string theory are carried out.
\end{abstract}

\section{The Hamiltonian Method}
The best way to get a feeling of how the Dirac Programme works is to study some
simple examples. Consider a set of non-relativistic particles, described by a
set of coordinates $q^i$, where $i$ labels the particle number as well as the
space coordinate, or think of the system as a set of particles in one space
dimension. More generally, the $q^i$ can be thought of as coordinates on a
manifold, which is called the {\em configuration space\/} $\conf$. The
classical motion of our system is given by a path in $\conf$, i.e.\ a function
$\q(t)$, where $t$ is the time and $\q=(\dots,q^i,\dots)$ denotes a point in
$\conf$. Such a path will be called a {\em time evolution}, or simply an {\em
evolution}.

The physical dynamics of our system is described by an action functional, which
assigns a real number to each evolution. In ``real life'' only those evolutions
are realized for which this action becomes extremal. What we need for our
purpose is that the action is given as a time integral of a {\em Lagrangian},
which is a function of the $q^i$ and their {\em velocities\/} $\dt q^i$, i.e.
\beq[action]
 \act[\q(t)] = \intdt  \lag(\q(t),\dt\q(t)).
\eeq
In principle, we could also consider Lagrangians which depend on higher
derivatives of $\q$ or explicitly on $t$. For the former, it is possible to
introduce additional variables, such that they can always be rewritten in the
form \eref{action}. If the Lagrangian depends on $t$ explicitly, the programme
to be presented below works in the same way, except that certain functions to
be introduced will become time-dependent as well. However, the most interesting
examples to which we want to apply the programme are those where not only the
Lagrangian is independent of $t$, but where even $t$ does not denote the
physical time (so a $t$-dependent Lagrangian wouldn't even make sense). But let
us come to this later on and stick to $t$ as the physical time for the moment.
I'll leave it as an exercise to modify what follows such that it can be applied
to time-dependent Lagrangians. Hence,~\eref{action} will be the most general
action to be considered here.

To be more precise, $\lag$ is actually a function on the tangent bundle of
$\conf$, which means that it takes a point $\q\in\conf$ and a vector $\v\in
T_\q\conf$ at $\q$ as arguments. Only after choosing a particular path $\q(t)$,
we can set $\v(t)=\dt \q(t)$ and evaluate the integral \eref{action}. So, we
shall in the following avoid the somewhat sloppy notation $\lag(\q,\dt\q)$ and
write $\lag(\q,\v)$ instead. A typical Lagrangian for a set of non-relativistic
particles is then given by
\beq[lag-nrel]
 \lag(\q,\v) = \ft12 M_{ij} v^i v^j - \pot(\q),
\eeq
where $M_{ij}$ is some ``mass matrix'' (e.g. $M_{ij} = m \delta_{ij}$ for a
collection of particles with the same mass or one particle in some higher
dimensional space), and the Einstein convention is used to indicate summation
over indices. We know how to derive the equations of motions from the
Lagrangian. By varying the action with respect to the path $\q(t)$, we find
that the time evolution has to satisfy the Euler Lagrange equations
\beq[euler-lagrange]
\dd/t/ \deldel \lag / v^i /(\q(t),\dt\q(t)) =
       \deldel \lag / q^i /(\q(t),\dt\q(t)).
\eeq
In general, these are second order differential equations for the function
$\q(t)$, and a time evolution which satisfies these equations will be called a
{\em solution}.

The classical part of the Dirac programme is essentially to transform the Euler
Lagrange equations into a standard form which can be used to classify their
solutions. First of all, as for any set of differential equations, they can be
rendered first order by introducing some auxiliary variables. In this case, we
need one additional variable $p_i$ for each $q^i$, and the most reasonable way
to do this is to choose $p_i$ to be the term under the time derivative on the
left hand side of \eref{euler-lagrange}, which gives the set of first order
equations
\beq[first-order]
        p_i(t)=\deldel \lag / v^i /(\q(t),\dt\q(t)), \qquad
    \dt p_i(t)=\deldel \lag / q^i /(\q(t),\dt\q(t)).
\eeq
If the configuration space is not a vector space, then the {\em momenta\/}
$p_i$ have to form a covector at $\q$, i.e.\ an element of the cotangent space
$T^*_\q\conf$. The time evolution is now given as a path in the cotangent
bundle $T^*\conf$, which is called the {\em phase space\/} $\phas$. spanned by
the coordinates $q^i$ and $p_i$. A point in $\phas$ will be called a {\em
state}. For our non-relativistic particles we find
\beq
    p_i(t) = M_{ij} \dt q^j(t) , \qquad
   \dt p_i(t) = - \deldel \pot / q^i /(\q(t)) .
\eeq
If the mass matrix is diagonal, or if there is at least an inverse satisfying
$M^{ij} M_{jk}=\delta^i\_k$, we get
\beq[ex-evol]
   \dt q^i(t) = M^{ij} p_j(t) , \qquad
   \dt p_i(t) = - \deldel \pot / q^i /(\q(t)) .
\eeq
This is a set of first order differential equation in the standard form
$\dt\x(t) = F(\x(t))$, where $\x(t)\in\phas$ is the state evolving in time. We
can start with any initial state $\x_0=(\q_0,\p_0)$ at, say, $t=0$, and get a
corresponding solution of the equations of motion by integrating
\eref{ex-evol}. Ignoring the problem of singularities, there is a one-to-one
correspondence between points in phase space and solutions to the
Euler-Lagrange equations.

Let us now come back to the general equations of motion \eref{first-order}. Can
we rewrite them in the standard form $\dt\x(t)=F(\x(t))$ as well? An elegant
way to answer this question is to get back to the action principle
\eref{action}. In addition to the momenta, we also introduce velocities $v^i$
as auxiliary variables, and define a new action
\beq[first-act]
 \acth[\q,\p,\v] = \intdt \dt q^i p_i - p_i v^i + \lag(\q,\v),
\eeq
where $\q(t),\p(t),\v(t)$ are to be treated as independent functions of the
time $t$. Variation with respect to $\p$ and $\v$ gives
\beq[first-vary]
   v^i = \dt q^i , \qquad
   p_i = \deldel \lag / v^i / (\q,\v).
\eeq
Reinserting this into the action takes us back to \eref{action}, together with
the correct definition for the momenta. So the new first order variational
principle is equivalent to the former. To obtain a set of first order
differential equations in standard form out of \eref{first-act}, all we have to
do is to get rid of the $\v$ variables therein. They are not very interesting
anyway, because, for any solution, they are nothing but the time derivative of
$\q$. Let's try the following. On the phase space, we introduce the {\em
Hamiltonian\/}
\beq[ham-def]
  \ham(\q,\p) = \Ext_\v \big(\, p_i v^i - \lag(\p,\v)\big),
\eeq
where $\Ext_\v$ denotes the extremum of the expression to the right, with
respect to the variables $\v$. For the moment we assume that there is a unique
such function. Then, we can go over to a new action principle which is entirely
defined on $\phas$ and given by
\beq[first-act-2]
  \acth[\q,\p] = \intdt \dt q^i p_i - \ham(\q,\p).
\eeq
The corresponding equations of motion are the Hamilton equations
\beq[hjeq]
   \dt q^i(t) =  \deldel \ham / p_i / (\q(t),\p(t)), \qquad
   \dt p_i(t) =  - \deldel \ham / q^i / (\q(t),\p(t)).
\eeq
They provide a set of first order differential equations in the standard form,
which describe the time evolution of the state, and they are equivalent to the
Euler Lagrange equations. To simplify the notation, we introduce an important
structure on the phase space, the {\em Poisson bracket}. To motivate its
definition, take a function $F(\q,\p)$ on $\phas$ and consider its time
evolution, i.e.\ the value of $F(t)=F(\q(t),\p(t))$. Using \eref{hjeq} we find
\beq[F-dt]
  \dt F(t) = \deldel F / q^i / \deldel \ham / p_i / -
          \deldel \ham / q^i / \deldel F / p_i /.
\eeq
Note that $q^i$ and $p_j$ are just some special functions on $\phas$, so
\eref{hjeq} can be expressed in this way as well. The right hand side is
another function on $\phas$, which has to be evaluated at $\q(t),\p(t)$, and
this function is called the {\em Poisson bracket\/} of $F$ and $\ham$.
Generally, the Poisson bracket maps two phase space functions $F$ and $G$ onto
a third function
\beq[pois-def]
   \pois FG  =  \deldel F / q^i / \deldel G / p_i / -
                \deldel G / q^i / \deldel F / p_i / .
\eeq
It's basic properties are antisymmetry, linearity, the Leibnitz rule and the
Jacobi identity:
\beq[pois-prop]
   &&\pois FG = - \pois GF, \qquad
   \pois {F + G}K = \pois FK + \pois GK, \zl
   &&\pois {FG}K = \pois FKG + F \pois GK, \zl
   &&\pois {\pois FG} K + \pois {\pois KF} G + \pois {\pois GK} F = 0.
\eeq
I will not prove this here as all this should be well known. We can replace the
equations of motion by requiring that
\beq[F-evol]
  \dt F = \pois F\ham
\eeq
for every phase space function $F$. By inserting the special functions $q^i$
and $p_i$, we recover \eref{hjeq}, and due to \eref{F-dt} the converse is true
as well. If we are a little bit sloppy, we can also write
\beq[x-evol]
  \dt \x = \pois \x\ham,
\eeq
but here we have to interpret $\x$ as a {\em collection of coordinates\/}
rather than a {\em point} on $\phas$, as the Poisson bracket takes functions on
$\phas$ as arguments and not points. Nevertheless it is a useful notation and
should not lead to confusion. As a result, the classical dynamics of our system
is now completely determined by two basic objects on the phase space: the
Hamilton function and the Poisson bracket. But all this was based on the fact
that the Hamiltonian can be defined by \eref{ham-def}. A system which can be
treated in this way will be called {\em unconstrained} (we will see shortly
why). In the context of the Dirac programme these are not of much interest, as
everything what follows, at least at the classical level, will be rather
trivial. Let us nevertheless stick to such a system for a while, just to
introduce some more notions which will be useful below.

A function on $\phas$ generally denotes a quantity that (in principle) can be
``observed'', like the positions of the particles, or the distances between
them or whatever. Let us call them {\em observables}. With the Poisson bracket
the set of observables becomes a Lie algebra. If $F$ is an observable, then
\eref{F-evol} tells us that there is another observable associated with the
time derivative of $F$. An observable $Q$ whose Poisson bracket with $\ham$
vanishes is called a {\em constant of motion\/} or {\em conserved charge}, as
whatever the state of the system is the value of $Q(t)$ will be constant. The
constant functions on $\phas$ are the trivial conserved charges, and if $Q$ and
$P$ are two conserved charges, then so are $Q+P$, $QP$, and also $\pois QP$.
The constants of motion form a subalgebra of the observables. The Hamiltonian
is always a conserved charge, which is usually called the energy.

As is also well known from classical mechanics, conserved charges are
associated with symmetries via the Noether theorem. How does this show up here?
As the conserved charges form a Lie algebra, we should expect that these are
the infinitesimal generators of symmetry transformations. Let $Q$ be some
conserved charge, then the symmetry transformation generated by $Q$ is
\beq[sym-trans]
  \delta \x = \pois \x Q,
\eeq
where the bracket is to be understood as in \eref{x-evol}. By using the Jacobi
identity and the fact that $\{Q,\ham\}=0$, we see that with $\x(t)$,
$\x(t)+\delta\x(t)$ is a solution as well, provided that we choose a
time-independent $Q$. You can also check that the commutator of two
infinitesimal symmetry transformations gives the transformation generated by
the Poisson bracket of the two generating functions, so the algebra of
conserved charges is in fact the Lie algebra associated with the symmetry group
of the system. If we take $\ham$ as a generator of a symmetry, then
\eref{sym-trans} formally coincides with \eref{x-evol}. This  means that the
symmetry transformation induced by $\ham$ is the time translation, which is
always a symmetry as we started from a time-independent Lagrangian.

How can we now apply this ``Hamiltonian method'' to a more general Lagrangian?
Suppose that the mass matrix in our example above becomes singular.
Then there will in general be no extremum of the function appearing in
\eref{ham-def}. We have to find another way to eliminate the $\v$ variables
from \eref{first-act}. Let's have a look at the equation of motion for $\v$,
which reads
\beq[v-var]
  p_i = \deldel \lag / v^i / (\q,\v).
\eeq
It does not contain any time derivatives, and of course it is just the
condition we have to solve to find the extremum in \eref{ham-def}. Now suppose
it cannot be solved for a given state $(\q,\p)$. This means that the
Hamiltonian cannot be defined at this point of the phase space, but it also
means that there is no solution to the equations of motion passing through this
point in $\phas$. So perhaps we don't need the Hamiltonian there, as we will
never get there. On the other hand, if \eref{v-var} can be solved, there {\em
is} an extremum in \eref{ham-def}, and provided that the Lagrangian is not too
ill-natured\footnote{In general, there will be not only one value $\v$ where
the function in \eref{ham-def} becomes extremal, but in almost all ``physical''
cases the set of these values is connected, so that the actual extremum is
constant. This requires the Lagrangian to be in some sense ``regular'', e.g.\
at most quadratic in the velocities. In all our examples this will be the case,
but there are funny systems where this is not true. Take, e.g., $\lag=v^3$ and
consider piecewise smooth time evolutions. Then solve the extremum condition
for the action \eref{action} and try to reproduce these solutions with the
Hamiltonian method!}, $\ham(\p,\q)$ is well-defined.
So we can only define the Hamiltonian on a subspace of $\phas$ which is given
by the image of the {\em momentum map\/} \eref{v-var}, as only for those states
we can find the extremum required for $\ham$. Again assuming that the
Lagrangian is not to weird, this image can be described in terms of a set of
equations $\con_\alpha(\q,\p)=0$, which we shall call the {\em primary
constraints}. The subspace defined in this way is called the {\em primary
constraint surface} (of course, the name suggests that there will be more
constraints later on).

Knowing that all possible solution entirely lie inside the primary constraint
surface, and that $\ham$ is well defined thereon, we can again replace the
action principle by \eref{first-act-2}, but now with the restriction that the
path has to lie inside the primary constraint surface. There is a general
method to deal with such ``constrained variational problems''. We can make the
explicit restriction implicit by adding Lagrange multiplier terms to the action
and define
\beq[act-ham-u]
  \acth[\q,\p,\u] = \intdt \dt q^i p_i - \phm(\q,\p)
                      - u^\alpha \con_\alpha(\q,\p),
\eeq
where $\phm$ has to coincide with $\ham$ on the primary constraint surface, but
is otherwise arbitrary (choosing a different extension of $\ham$ corresponds to
a transformation of the $\u$ variables). Defining the combination
\beq[ham-con]
  \ham (\q,\p) = \phm(\q,\p) + u^\alpha \con_\alpha(\q,\p)
\eeq
to be the Hamiltonian, the complete set of equations of motions become
\beq[con-evol]
   \con_\alpha = 0 , \qquad
   \dt q^i = \deldel \ham / p_i / , \qquad
   \dt p_i = \deldel \ham / q^i / .
\eeq
They already look similar to \eref{hjeq}, but there are some crucial
differences. So, for example, the auxiliary variables $u^\alpha$ are still in
there, which seem to have no physical meaning. In fact, they parametrize the
Hamiltonian, so we do not only have one Hamiltonian but a whole set of them.
How do we know which is the right one, or are we free to choose it as we like?
The $u^\alpha$ are quite implicit in \eref{con-evol}, so it is not obvious what
their meaning is. Before coming to these questions in general, let us consider
some examples.

A singular mass matrix occurs whenever there are coordinates whose velocities
do not appear in the Lagrangian. Such variables typically appear as Lagrange
multipliers, and are in classical mechanics used to impose constraints onto a
system, just as in our auxiliary action above. But now we will have such terms
already in the original action. A nice example is the following. The
configuration space is three dimensional, with coordinates $q_1,q_2,q_3$, and
the Lagrangian is
\beq[c-lag]
\lag = \ft12 m \big( v_1\^2 + v_2\^2 \big)
               - \ft12 q_3 ( q_1\^2 + q_2\^2 - r^2 ).
\eeq
It should not be too difficult to see that this describes a particle of mass
$m$ which moves on a circle of radius $r$ in a two dimensional plane spanned by
$q_1,q_2$, with $q_3$ being the force necessary to make the particle stay on
the circle. The general solution to the Euler Lagrange equations are
\beq[c-sol]
 q_1(t)= r \cos(\omega t + \varphi), \qquad
 q_2(t)= r \sin(\omega t + \varphi), \qquad
 q_3(t)= m \omega^2,
\eeq
parametrized by $\omega,\varphi\in\RR$. The energy of the particle is $\ft12 m
r^2 \omega^2$, which can be expressed in terms of the configurations variables
as $\ft12 r^2 q_3$. So we expect this to be the Hamiltonian. Let us just keep
this in mind, and try to reproduce the solution by applying the Hamiltonian
method. By differentiating the Lagrangian we find the following momentum map:
\beq[c-mom]
  p_1 = \deldel \lag / v_1 / =  m v_1, \qquad
  p_2 = \deldel \lag / v_2 / =  m v_2, \qquad
  p_3 = \deldel \lag / v_3 / =  0.
\eeq
The image of this map is obviously given by the subspace $p_3=0$ of the phase
space, so our primary constraint is
\beq[const-1]
  \con_1(\q,\p) = p_3.
\eeq
It is not difficult to find the Hamiltonian by solving \eref{c-mom} for $\v$
and insert the solution into $p_iv_i-\lag$. Note that $v_3$ remains completely
arbitrary, but the {\em value\/} of the extremum does not depend on $v_3$, as
$p_3=0$ on the constraint surface. The total Hamiltonian analogous to
\eref{ham-con} then becomes
\beq
  \ham(\q,\p) = \ft12 m^{-1} ( p_1 \^2 + p_2\^2 )
                + \ft12 q_3 \, ( q_1\^2 + q_2\^2 - r^2 ) + u \, \con_1 ,
\eeq
and the time evolution equations read
\beq[c-evol]
  &&\dt q_1 =  \deldel \ham / p_1 / =  m^{-1} p_1 , \qquad
    \dt p_1 =  -\deldel \ham / q_1 / = - q_3 q_1 , \zl
  &&\dt q_2 =  \deldel \ham / p_2 / =  m^{-1} p_2 , \qquad
    \dt p_2 =  -\deldel \ham / q_2 / = - q_3 q_2 , \zl
  &&\dt q_3 =  \deldel \ham / p_3 / =  u , \qquad
    \dt p_3 =  -\deldel \ham / q_3 / = - \ft12( q_1\^2 + q_2\^2 - r^2 ) .
\eeq
Let's try to integrate these equations, starting with a state on the constraint
surface, i.e.\ with $p_3=0$. This doesn't work quite well, because the last
equation immediately tells us that $p_3$ will not stay zero, unless $q_1\^2 +
q_2\^2 = r^2$. Hence, it is not enough to start with a state on the primary
constraint surface, as this will in general evolve away from it. There are more
restrictions on the initial condition than just $p_3=0$. Its time derivative
must vanish as well, which gives us another constraint
\beq
  \con_2 = \ft12( q_1\^2 + q_2\^2 - r^2 ) .
\eeq
Note that this is implicit in \eref{c-evol}, so it is actually not a new
equation of motion, but its puts a {\em new\/} restriction on the initial
conditions, which have to be fulfilled before we start to integrate the time
evolution. We shall call this a {\em secondary constraint}. In other words, it
singles out jet another subspace of $\phas$. Of course, this is just the
restriction that our particle moves on a circle with radius $r$. So we have to
add this to the constraints.

Are we finished now? Starting with a state $(\q,\p)$ satisfying both
constraints, will we stay on the constraint surface for all times? To find this
out, all we have to do is to compute the time evolution of the phase space
function $\con_2$, which gives
\beq
  \dt \con_2 = \pois{\con_2}{\ham} =
         m^{-1} ( p_1 q_1 + p_2 q_2 ).
\eeq
So we need another constraint
\beq
  \con_3 =   p_1 q_1 + p_2 q_2 .
\eeq
More?
\beq
  \dt \con_3 = \pois{\con_3}{\ham} =
      - q_3 \big(q_1\^2 + q_2\^2\big) + m^{-1} \big(p_1\^2 + p_2\^2\big) .
\eeq
Again, a new constraint, but we can simplify it a bit using $\con_2$:
\beq
  \con_4 =  q_3 - m^{-1} r^{-2} (p_1\^2 + p_2\^2)
\eeq
Same procedure again:
\beq
  \dt \con_4 = \pois{\con_4}{\ham} =
          u + 2 m^{-1} r^{-2} q_3 (p_1 q_1 + p_2 q_2).
\eeq
This is something new. It does not give a new constraint, but a restriction on
$u$. As it is sufficient that the right hand side vanishes on the constraint
surface (now including {\em all\/} the constrains collected so far), we only
have to require $u=0$ on that surface, but arbitrary outside. So, our original
idea that $u$ might enter the Hamiltonian as a free parameter can not be
confirmed. It is at least fixed on the constraint surface.

But now we are through. We found all constraints, four all together, and once
we start with a state solving them, we will stay on the constraint surface for
all times. It is not difficult to see that the general solution to {\em all\/}
constraints is
\beq[c-con-sol]
  &&q_1 = r \cos \varphi , \quad
  q_2 = r \sin \varphi , \quad
  q_3 = m \omega^2, \zl
  &&p_1 = - m \omega r \sin \varphi, \quad
  p_2 =  m \omega r \cos \varphi, \quad
  p_3 = 0,
\eeq
with $\varphi,\omega$ two real parameters, and that the time evolution is given
by replacing $\varphi \mapsto \varphi + \omega t$, leading back to
\eref{c-sol}.
So, this was a quite instructive example to show how secondary constraints are
derived, leading to a smaller constraint surface, which finally became two
dimensional. We shall call this the {\em constraint surface\/} $\phys$. It can
also be called the {\em physical phase space}, but sometimes this notion is
used differently, so let us stick to the first one. In our example, the
constraint surface looks like the phase space of a particle on a circle, with
configuration variable $\varphi$ and momentum $m\omega r$. But the Hamiltonian
is not simply the energy, which was found to be $\ft12 r^2q_3$. However, if we
look at it more closely, we find that
\beq[c-ham]
  \ham = \ft 12 r^2 \, q_3  + u_\alpha\, \con_\alpha, \txt{with}
   u_1 = 0 , \quad
  u_2 = q_3 , \quad
  u_3 = 0, \quad
  u_4 = - \ft12 r^2.
\eeq
We see that, on the constraint surface, the Hamiltonian indeed gives the
energy, but there are extra contributions proportional to the constraints. What
is the reason for this? To derive the evolution equations in \eref{c-evol}, we
need the derivatives of $\ham$ rather than $\ham$ itself, so it is not
sufficient two know $\ham$ on the constraint surface, but we also need to know
its first derivatives. We do not need to know its second derivatives or the
value of $\ham$ ``far away'' from the constraint surface. This means that we
are free to add anything to $\ham$ which vanishes {\em and\/} whose gradient
vanishes on the constraint surface, i.e. which is proportional to the square of
the constraints. But this again means that we are free to add anything that
vanishes on the constraint surface to the $u$ parameters. To express such a
condition in a convenient way, we introduce the following notations. Two phase
space functions $F$ and $G$ are defined to be {\em weakly equal\/}, written as
$F\weak G$, if they coincide on the constraint surface. The conditions to be
imposed on the $u$'s can then be written as
\beq
  u_1 \weak 0 , \qquad
  u_2 \weak q_3 , \qquad
  u_3 \weak 0, \qquad
  u_4 \weak - \ft12 r^2.
\eeq
With weak equalities we have to be careful, because they are not compatible
with the Poisson bracket. If we have $F\weak G$, this does not imply
$\{F,K\}\weak\{G,K\}$ in general. We are not allowed to use weak equalities
inside the brackets. The result is that the Hamiltonian, as a phase space
function, is only fixed up to terms of second order in the constraints, but
this is all we need as only first derivatives of $\ham$ on the constraint
surface appear in the Hamilton equations.

Let us consider jet another example, which in a sense will also turn out to be
``unnecessarily complicated'' because of ``unconventional'' variables, but it
behaves quite differently from the former is explains another important issue
concerning constrained systems. This time, the configuration space is two
dimensional and the Lagrangian is
\beq
   \lag = \ft12 m \, (v_1+v_2)^2 - \pot(q_1+q_2).
\eeq
It looks a bit silly, you will ask why shouldn't we take $q=q_1+q_2$ as a new
variable, then the second will drop out and we are left with a single particle
in one dimension. But this system will provide us with the simplest example of
a {\em gauge theory}. The momenta are
\beq
  p_1 = m (v_1 + v_2 ), \qquad
   p_2 = m (v_1 + v_2 ),
\eeq
which gives the primary constraint
\beq
  \con = p_1 - p_2.
\eeq
Again, the Hamiltonian contains one free parameter and is easily evaluated to
\beq
 \ham(\q,\p) =  \ft12 m^{-1} p_1\^2 + \pot(q_1+q_2) + u \, \con.
\eeq
Note that we could also put $p_2$ into the first term, which gives the same
function on the constraint surface, but this can be compensated by replacing
$u$ with $u-\ft12 m (p_1 + p_2)$. We also find that $\{\con,\ham\}=0$, so there
are no more constraints, nor is there any restriction on $u$. The evolution
equations become
\beq
  \dt q_1 = m^{-1} p_1 + u , \quad
  \dt q_2 = -u  , \quad
  \dt p_1 = -\pot'(q_1+q_2), \quad
  \dt p_2 = -\pot'(q_1+q_2),
\eeq
showing explicitly that the constraint is conserved. We see that, on the
constraint surface, the Hamiltonian gives the energy of a particle at $q_1+q_2$
with momentum $p_1 = p_2 = m ( \dt q_1 + \dt q_2)$. The physical interpretation
of this system is as follows. There is a particle in one dimension whose
position is $q_1+q_2$, but it has some internal degree of freedom which can be
taken to be the difference $q_1-q_2$. The time evolution of this degree of
freedom is completely arbitrary. Moreover, we see what the role of the
constraint in this case is. It is the generator of a symmetry transformation,
which is given by
\beq
  &&\delta q_1 = u \,\pois{q_1}{\con} = u, \qquad
  \delta q_2 = u \,\pois{q_2}{\con} = -u , \zl
  &&\delta p_1 = u \,\pois{p_1}{\con} = 0, \qquad
  \delta p_2 = u \,\pois{p_2}{\con} = 0.
\eeq
It leaves the ``physical'' quantities, the position $q_1+q_2$ and the momentum
invariant, but changes the internal degree of freedom. The time evolution
consists of a ``physical'' evolution of the position and momentum, plus a
transformation generated by the constraint, with an arbitrarily time-dependent
parameter. Starting from the same initial conditions, the state can evolve into
different final states, as we are free to choose $u$. If we want physics to be
deterministic, we have to assume that this degree of freedom is unphysical, and
cannot be measured. Otherwise we would not find classical physics to be
deterministic. Two states, which can evolve out of a single state in this way
must be declared to be physically equivalent, and a transformation that takes
such states into each other is called a {\em gauge transformation}.
It is important to note that all this does not come out because we didn't
impose enough initial conditions. Perhaps, if we, in addition to the
velocities, also fix the acceleration, then the evolution might become unique
again. But this is not true, which follows from a theorem about differential
equations, but which can also be seen quite directly. Assume we have a solution
which was generated by some choice for the $u$ parameter. Then, change this
parameter, but only for times $t>t_0$. The new solution will be identical to
the old one for $t<t_0$, but different thereafter. Hence, there is no chance to
fix the time evolution by giving more initial conditions.

As a result, we found that there are different kinds of constraints. In the
first example there were no gauge degrees of freedom, and the Hamiltonian was
unique (up to second order terms in the constraints). In the second example, it
was not unique, but contained one free parameter (of course, here we are also
free to add anything proportional to the square of the constraint, so we can
make $u$ any time-dependent phase space function). Can we somehow see this
difference by expecting the constraints themselves? In fact, a general
criterion can be found to distinguish these two {\em classes\/} of constraints,
and to work this out let us come back to the general theory.

What we had so far was that the equations of motion can be derived from a
Hamiltonian which takes the form
\beq[prim-ham]
  \ham(\q,\p) = \phm(\q,\p) + u^\alpha \, \con_\alpha(\q,\p),
\eeq
where $\con_\alpha$ are the primary constraints derived from \eref{v-var}. The
$u$ were introduced as a set of free parameters, but to be a bit more general
we can also allow them to be a set of phase space functions {\em containing}
some free parameters (possibly we are forced to do so, see \eref{c-ham}). If we
want to use the general formula $\dt \x = \{\x,\ham\}$ to obtain the time
evolution, we first have to find out which initial states are allowed. From the
example we learned that this can be done be requiring that, beside the primary
constraints, also their time derivatives vanish. It is enough for them to
vanish weakly, because this already ensures that the state remains on the
constraint surface, so we have to consider the conditions
\beq[sec-const]
  \dt \con_\alpha = \pois{\con_\alpha}{\ham} \weak
   \pois{\con_\alpha}{\phm} + u^\beta \pois{\con_\alpha}{\con_\beta}
     \weak 0.
\eeq
If the $u$'s are phase space functions, we have to take this into account when
we compute brackets with $\ham$. However, the relevant part of the result is
always proportional to some constraint, so it vanishes weakly, and we made use
of this in \eref{sec-const}. We can split this set of equations into a set of
secondary constraints, and some equations restricting the $u$'s, which
effectively reduce the number of free parameters. Of course, the newly found
constraints, also denoted by $\con_\alpha$ with an extended range for $\alpha$,
have to be conserved under the time evolution as well. So we again have to
evaluate \eref{sec-const} for them, possible getting more constraints and so
on, until we either arrive at a contradiction or the new conditions are
trivial. In the first case, there are no solutions to the equations of motion,
the theory is rather uninteresting. In the second case, we end up with a full
set of constraints
\beq[tot-con]
 \con_\alpha(\q,\p)=0,
\eeq
which define the constraint surface $\phys$ and the weak equality relation. The
Hamiltonian takes the form
\beq[tot-ham]
 \ham(\q,\p) = \phm(\q,\p) + u^\alpha(\q,\p) \, \con_\alpha(\q,\p),
\eeq
where the $u^\alpha$ may contain some free parameters, and are otherwise fixed
weakly only. Note that there is a non-trivial step from \eref{prim-ham} to
\eref{tot-ham}, as now we are summing over {\em all\/} constraints, not just
the primary ones. This is justified because we can now modify the Hamiltonian
anywhere except at the constraint surface, which is in general smaller than the
primary constraint surface. The $u$'s corresponding to secondary constraints
will however be fixed, because free parameters only appear in front of the
primary constraints (if $\phm$ is taken to be the same function as in
\eref{prim-ham}, we simply have $u^\beta\weak0$ for the secondary ones). The
question that remains to be answered is whether there are any free parameters
left after this procedure. Can we perhaps localize them without solving the
equations \eref{sec-const} explicitly? We would then also know which
constraints are those that generate gauge transformations and which are not.

After finding all the constraints, we can once again have a look at the
equations for the $u^\alpha$ parameters. They were all of the form
\eref{sec-const}. If we now take all constraints and the Hamiltonian from
\eref{tot-ham}, then the $u$'s have to satisfy the following set of weak
equations
\beq[u-eq]
  \pois{\phm}{\con_\alpha} + u^\beta \, \pois{\con_\alpha}{\con_\beta}
   \weak 0,
\eeq
which will in general be over-complete as part of it has already been solved to
derive the secondary constraints, but nevertheless it forms a full set of
equations for the $u$'s. Actually, only the $u$'s corresponding to the {\em
primary\/} constraints are entering these equations as variables, but we can
take them as equations for the secondary $u$'s as well for the moment, and look
for the general solution (we know that there is a solution with $u^\beta\weak0$
for the secondary $u$'s, but there might be more and below we will see that
these turn out to be useful as well). Whether or not there are free parameters
in the $u$'s now crucially depends on the antisymmetric square matrix
\beq
  \pmat_{\alpha\beta} = \pois{\con_\alpha}{\con_\beta}.
\eeq
If it is invertible, we can solve \eref{u-eq}, giving us a (weakly) unique
solution for all $u$'s, and thus a unique Hamiltonian and a unique time
evolution. By computing the relevant $4\times4$ matrix you can check that this
is the case for the particle on the circle.
But suppose the matrix is not invertible. Then there are some zero
eigenvectors. As we are free to make linear transformations on the constraints,
we can assume that there is a subset of constraints $\fcl_a$, whose brackets
with {\em all\/} constraints weakly vanishes, i.e.\
\beq
  \pois{\fcl_a}{\con_\alpha} \weak 0.
\eeq
It is obvious that then \eref{u-eq} does not impose any restriction on the
corresponding $u^a$. A phase space function with the property that it has
weakly vanishing Poisson brackets with all constraints, is called a {\em first
class\/} function (as an example, the Hamiltonian is always first class). So
the $\fcl_a$ form the set of {\em first class constraints}. On the other hand,
if we denote the remaining constraints by $\scl_\ba$, then the sub-matrix
$\pmat_{\ba\bb} = \pois{\scl_\ba}{\scl_\bb}$ is invertible (if it was not, then
there must be a linear combination of the $\scl_\ba$ which is first class). As
a result, the $u^\ba$ corresponding to these constraints are (weakly) fixed.
Let's call the $\scl_\ba$ {\em second class constraints}. The whole set of
constraints is now split into first and second class constraints, where the
first class constraints form a linear subset, and the second class constraints
are fixed up to linear combinations of first class constraints
only.\footnote{Actually this split can only be made separately at each point on
the constraint surface. However, if we again assume that the Lagrangian is
somehow regular, we can do the split globally. This requires, e.g., that the
(weak) rank of the matrix $\pmat_{\alpha\beta}$ is constant, which will be the
case in all our examples. There are, however, some interesting theories,
especially gravity theories with singular metrics, which do not have this
property.}

{}From the example we already learned that gauge transformations are generated
in the same way as the symmetry transformations considered in \eref{sym-trans},
with the conserved charge replaced by the constraint. They show up in the
Hamiltonian as a term proportional to the constraint, with a free parameter in
front. If $F$ is some phase space function, then
\beq[gauge-trans]
  \delta F = u^a \, \pois F{\fcl_a} \weak \pois F{u^a \, \fcl_a}
\eeq
gives the transformation of $F$ generated by the special linear combination
$u^a\fcl_a$. For our general Hamiltonian, the constraints appearing in this way
are exactly the {\em primary first class\/} constraints. So we conclude that
these are the generators of gauge transformations. However, if we look at the
Hamiltonian and the constraints in their final form, the distinction between
primary and secondary constrains doesn't show up anywhere except in the fact
that {\em originally\/} only the primary ones had free $u$ parameters in front.
But to extract the first class constraints we had to find the zero eigenvectors
of the matrix $\pmat_{\alpha\beta}$, and this might involve a mixing of primary
and secondary constraints, so it is actually not clear whether one can define
{\em primary first class\/} constraints at all. And it would be much nicer if
we could forget about the distinction between primary and secondary
constraints, as it only shows up in the {\em derivation\/} of the constraints
and not in their final form (in contrast to the distinction between first and
second class constraints). So let us prove that the secondary first class
constraints generate gauge transformations as well.

For this proof, let as for the moment assume that there are no second class
constraints, then we have to prove that all constraints generate gauge
transformations. Otherwise, the proof goes through in the same way, it just
becomes technically more complicated.\footnote{Alternatively, one can use the
Dirac brackets to be introduced below and do the proof formally is given here.}
We know that every secondary constraint $\con$ can be written as a linear
combination of the form
\beq[gauge-proof]
   \con = c^\alpha \, \big(\pois{\con_\alpha}{\phm} +
              u^\beta \,  \pois{\con_\alpha}{\con_\beta} \big),
\eeq
where the sum runs over only those $\alpha$'s that correspond to constraints
discovered ``earlier'' than $\con$ (including the primary ones). It was exactly
that way the secondary constraints were derived, namely be eliminating the
$u$'s from \eref{sec-const}. If we assume that all constraints appearing on the
right hand side of \eref{gauge-proof} are generators of gauge transformations,
then so is $\con$, for the following reason. If two constraints generate a
gauge transformations, then their Poisson bracket does as well, because the
transformation generated is just the commutator of both, and any combination of
gauge transformation is gain a gauge transformations. Moreover, the commutator
of a gauge transformation with the time evolution is also a gauge
transformation. Whether we first make a gauge transformation and then evolve
our system, or the other way around, that does not make any difference. Hence,
the bracket of a constraint with the Hamiltonian is again a generator of a
gauge transformation. It now follows by induction that all constraints generate
gauge transformations, as we know that the primary ones do, and all others are
successively given by \eref{gauge-proof}.

Now comes a crucial step. If we are only interested in the physical state of
our system, we are free to make gauge transformations at any time, so we are
free to add any linear combinations of first class constraints to $\ham$. The
parameters in front of these constraints can be any phase space function which
can also depend on time. So, finally we end up with the Hamiltonian taking the
form
\beq[ext-ham]
  \ham (\q,\p,t) = \phm(\q,\p) + u^a(\q,\p,t) \, \fcl_a(\q,\p).
\eeq
Here, the terms containing the second class constraints have been included into
$\phm$ as they do not contain any free parameters. I should emphasize, that
this step is entirely motivated by {\em physics} and takes us away from the
purely mathematical question of finding time evolutions extremizing the action.
This Hamiltonian no longer generates solutions to the original Euler Lagrange
equation. But physically a time evolution obtained by integrating the Hamilton
equations starting from a state on the constraint surface will always be
gauge-equivalent to some solutions of the Euler Lagrange equations. Finally,
the complete set of equations of motions split into the first and second class
constraints
\beq[all-con]
  \fcl_a(\q,\p) = 0, \qquad
  \scl_\ba(\q,\p) = 0,
\eeq
and the Hamilton equations
\beq[tot-evol]
   \dt q^i = \deldel \ham / p_i / , \qquad
   \dt p_i = \deldel \ham / q^i / .
\eeq
Using the bracket notation, they read
\beq[x-tot-evol]
   \dt \x = \pois{\x}{\ham} \weak
            \pois{\x}{\phm} + u^a \, \pois{\x}{\fcl_a},
\eeq
showing how the time evolution is split into a ``physical'' evolution and a
gauge transformation. Notice, however, that here we are again using the
somewhat sloppy notation, taking the bracket with the {\em point\/} $\x$, which
should actually be a phase space function.

This in a way completes the classical part of the Dirac programme, but let us
consider the second class constraints a bit further. We got a quite nice
understanding of the physical meaning of the first class constraints, but yet
we have no idea whether there is something similar for the second class ones.
{}From our particle moving on a circle we know that these can emerge from
exterior constraints imposed on the configuration variables in the Lagrangian.
Maybe they are just a kind of ``trivial'' restrictions which simply reduce the
actual ``size'' of the phase space but are not very interesting otherwise. We
could equally well solve the particle on the circle by choosing only one
(periodic) configuration variable $\varphi$ and we would end up with the same
solutions. But we won't end up with the same {\em Hamiltonian}. This is because
in \eref{c-ham} there are contributions proportional to the second class
constraints which will never show up in the simplified model. So, it seems to
be non-trivial to get rid of the extra phase space variables {\em after\/}
applying the Hamiltonian method. If we simply change coordinates such that two
of them are the angle $\varphi$ and some conjugate momentum, the second class
constraints do not drop out form the Hamiltonian, because they will still be
necessary to provide the correct equations of motion. One should expect that
$\varphi$ and $\pi=m\omega r$ are the relevant phase space coordinates that
also appear in the simplified model as conjugate variables. But observe that,
depending on how they are defined as functions of our phase space variables,
their Poisson bracket will in general not (and not even weakly) be
$\{\phi,\pi\}=1$.\footnote{Take $\phi=\arctan(-p_1/p_2)$ and
$\pi^2=p_1\^2+p_2\^2$, then their bracket even vanishes.}

Nevertheless there is a way to get rid of the second class constraints in the
Hamiltonian. The trick is to modify the Poisson bracket such that all
constraints become first class. If we fix a set of second class constraints
$\scl_\ba$, it is possible to define a modified bracket, the {\em Dirac
bracket}, which still gives all the correct equations of motion, but for them
we will have
\beq
  \dirac F{\scl_\ba} \weak 0
\eeq
for {\em every\/} phase space function $F$. As a result, terms proportional to
them appearing in the Hamiltonian will not contribute to the evolution
equations (not even generating a gauge transformation), and we can drop the
corresponding terms (actually: we are free to add terms proportional to the
$\scl_\ba$ to any function without weakly changing their brackets, especially
to the Hamiltonian).  To see how this can work, let us consider the most simple
set of second class constraints. Suppose we have a phase space spanned by $q^i$
and $p_i$, $i=1,\dots,n$, and there are two constraints $\scl_1=q^1$ and
$\scl_2=p_1$. These are obviously second class. They just tell us that the
first degree of freedom is of no interest at all, we should have left it apart
from the very beginning. What would be the Poisson bracket on the phase space
if we started without the first degree of freedom? Clearly, it would be
\beq
  \dirac FG = \sum_{i=2}^n\,\deldel F / q^i / \deldel G / p_i / -
                            \deldel G / q^i / \deldel F / p_i / .
\eeq
We just have to drop $i=1$ from the summation. We can express this in terms of
the full Poisson bracket and the constraints only. It reads
\beq
  \dirac FG = \pois FG  + \pois{F}{\scl_1} \pois{\scl_2}{G}
                        - \pois{F}{\scl_2} \pois{\scl_1}{G}.
\eeq
You can easily check that this is exactly the same as above. In a sense, the
two constraints here are ``conjugate'' phase space variables, and what the last
equation does is to subtract that part of the Poisson bracket that comes from
this pair of variables. You can also check that the new bracket vanishes
whenever one of its arguments is one of the constrains, and as we shall see it
has all the properties of the Poisson bracket given in \eref{pois-prop}. It
seems that what we have to do is to find a transformation of variables such
that the second class constraints become pairwise conjugate. This needs their
number to be even, will this always be the case? Remember that we had a matrix
\beq[sec-mat]
  \pmat_{\ba\bb} = \pois{\scl_\ba}{\scl_\bb},
\eeq
and the second class constraints were defined such that this is non-singular.
But an antisymmetric matrix must have even rank, so the number of second class
constraints will always be even. It is not clear how this problem can be solved
in general, because it is not always possible to find these pairs of variables.
 However, we actually only need to solve a ``linearized version'' of the
problem. That is the following. If $F$ is some phase space function, then we
want to have
\beq[sec-van]
  \dirac F{\scl_\ba} \weak 0,
\eeq
for all second class constraints. We make an ansatz
\beq[dir-def]
  \dirac FG = \pois FG
     - \pois{F}{\scl_\ba} \,\imat^{\ba\bb}\, \pois{\scl_\bb}{G},
\eeq
which is similar to what we had above. You can now simply solve the equation
\eref{sec-van} to find that $\imat^{\ba\bb}$ has to be the inverse of
$\pmat_{\ba\bb}$, i.e.
\beq
  \imat^{\ba\bb} \, \pmat_{\bb\bc} =
      \imat^{\ba\bb} \,\pois{\scl_\bb}{\scl_\bc} = \delta^\ba\_\bc.
\eeq
Actually we only need this equation to be satisfied weakly in order to get
\eref{sec-van}. But if we can solve it ``strongly'', then $\{F,\scl_\ba\}_*$
would also vanish strongly, which might be quite useful as we can then use the
second class constraints anywhere inside the brackets, however nested they are.
In most cases this can be established, but not in general, as the matrix
$\pmat_{\ba\bb}$ might become singular at some point outside the constraints
surface. However, its determinant is non-zero on the surface, so there is a
(finite) neighbourhood were it can be inverted to give $\imat^{\ba\bb}$. Then,
\eref{sec-van} holds on a whole neighbourhood of $\phys$, and beside
$\{F,\scl_\ba\}_*$ all its derivatives will vanish, and we can use $\scl_\ba=0$
as a strong equality inside brackets, as long as we {\em finally\/} evaluated
the result on the constraint surface. So, with the Dirac bracket we don't have
to worry about the second class constraints any more, they just become strong
identities for all practical purposes.

But yet we haven't proved that they do the job, i.e.\ that they really provide
the same equations of motions as the Poisson bracket. This is not very
difficult, as
\beq
  \dirac{F}{\ham} = \pois{F}{\ham} - \pois{F}{\scl_\ba} \, \imat^{\ba\bb} \,
                  \pois{\scl_\bb}{\ham} \weak \pois{F}{\ham} = \dt F,
\eeq
for any phase space function $F$, because the Poisson bracket of the
Hamiltonian with any constraint vanishes weakly. There remain some other things
to be shown, namely that the new bracket has the same properties
\eref{pois-prop} as the old one. Except for the Jacobi identity this is
immediately obvious. With the formalism derived so far, the only possibility to
prove the Jacobi identity is explicit calculation, which is rather involved. As
always in cases like this, a very simple and almost trivial proof exist if one
uses a different formalism, which makes more use of the geometric structure of
the phase space. So, let us skip the rather tedious proof for the Jacobi
identity here.

After introducing the Dirac bracket, we are now also free to add anything
proportional to second class constraints to the Hamiltonian, without affecting
the equations of motion. Hence, we can take it to be
\beq
  \ham(\q,\p,t) = \phm(\q,\p) + u^a(\q,\p,t) \, \fcl_a(\q,\p) +
                                v^\ba(\q,\p,t) \, \scl_\ba(\q,\p),
\eeq
but we can equally well set $v^\ba=0$. However, the freedom to choose $v$ means
that the ``allowed'' Hamiltonians are now given by {\em all\/} phase space
functions weakly equal to $\phm$. Hence, we only need to know the energy, i.e.\
$\ham$ on the constraint surface. Its extension can be chosen arbitrarily. The
Dirac bracket is designed such that it ``picks up'' the derivatives of $\ham$
tangent to the surface and in the ``normal'' directions corresponding the first
class constraints only, thereby generating the undeterministic gauge
transformtions, but ignoring the derivatives in the direction of the second
class constraints.

Regarding the ``classical'' Dirac programme, we have more or less reached the
goal. Let us sumarize what we have found. The constraints single out a subspace
of the phase space. A point in that subspace corresponds to a physical state
and can be inserted into the Hamilton equations as initial condition. The first
class constraints generate gauge transformations. Moreover, they provide a
representation of the gauge algebra, which is the Lie algebra of the gauge
group associated with the system. The Lie bracket is of course given by the
Poisson bracket:
\beq[gauge-alg]
  \pois{\fcl_a}{\fcl_b} = f_{ab}\^c \, \fcl_c.
\eeq
We cannot expect that the {\em structure functions\/} $f_{ab}\^c$ are
constants, in general they will be some phase space function. We can only try
to redefine the constraints such that the structure functions become constant.
If that doesn't work, it will be of some importance for the quantum theory. But
why is the Poisson bracket of two first class constraints again a linear
combination of first class constraints? We know that it vanishes weakly, as
this was the definition of first class phase space functions, so it is
certainly a combination of constraints. So we have to prove that the bracket of
two first class functions $F$ and $G$ again gives a first class function. In
fact, the bracket of $\{F,G\}$ with some constraint gives
\beq
  \pois{\pois FG}{ \con } = \pois{\pois F\con}{G}-\pois{\pois G\con}{F}.
\eeq
Now $F$ is first class, so $\{F,\con\}$ can be written as is a linear
combination of constraints, whose bracket with $G$ vanishes weakly, and vice
verse, so the whole right hand side vanishes weakly. As a result, only first
class constraints appear on the right hand side of \eref{gauge-alg}.

For an unconstrained system we introduced {\em observables\/} as phase space
functions that correspond to measurable quantities. What is the analogy for
constrained systems? First of all, a measurable quantity should be a function
on the constraint surface only, as a state outside will never be realized. We
can nevertheless deal with functions on the whole phase space as {\em
representations\/} of observables. We simply define two function representing
the same observable if they are weakly equal. But if there are gauge degrees of
freedom, we had to assume that certain quantities are not measurable, as they
have an indeterministic time evolution. Only those quantities are measurable,
which are invariant under gauge transformations. As gauge transformations are
generated by the first class constraints, a function is invariant if its
brackets with the first class constraints vanish weakly. Using the Dirac
brackets, we can equally well say that the brackets with all constraints have
to vanish weakly:
\beq[obs-def]
   \pois{ \obs }{ \fcl_a } \weak 0 \equiv
   \dirac{ \obs }{ \con_a } \weak 0.
\eeq
We can obviously add and multiply observables, but what about taking brackets?
We have to assure that the resulting bracket is independent of the
representations chosen for the observables. So, let $F_1\weak F_2$ be two
representations of the same observable, and $G$ another observable, then
$F_1-F_2=u^\alpha\con_\alpha$ is some linear combination of constraints. From
\eref{obs-def} we conclude that its Dirac bracket with $G$ vanishes weakly, so
that
\beq
  \dirac {F_1}G \weak \dirac{F_2}G.
\eeq
The Dirac bracket therefore gives a unique operation on observables, which are
equivalence classes of phase space functions. As for the unconstrained systems,
the observables form a Lie algebra, but the Lie bracket is given by the Dirac
bracket and not by the Poisson bracket. We can also define {\em conserved
charges\/} as those observables which are constant in time. In addition to
\eref{obs-def}, a conserved charge $Q$ has to satisfy
\beq[crg-def]
  \dirac Q {\phm} \weak 0.
\eeq
It generates symmetry transformations
\beq[sym-gen]
  \delta F = \dirac F Q,
\eeq
which map solutions onto new solutions, as can be shown in the same way as for
unconstrained systems. Actually only the weak version of \eref{sym-gen} is
relevant, as solutions always live on the constraint surface. It is however
important that the Dirac bracket appears: take $F$ to be a second class
constraint, then we must clearly have $\delta F\weak 0$, which is only
guaranteed by the Dirac bracket. Moreover, we see that if $F$ represents an
observable and $Q$ a conserved charge, the $\delta F$ is again an observable,
which is independent of the chosen representation. So the symmetry
transformation are well defined on the observables.

If we just use the definition, the constraints themselves also become conserved
charges: the brackets \eref{obs-def} and \eref{crg-def} both vanish weakly. And
they are indeed generators of symmetries, which are however ``unphysical'',
i.e.\ they do not change the physical state (whereas in general symmetry
transformation do change the physical state). But remember that we defined
observables to be {\em equivalence classes\/} of phase space functions with
respect to $\weak$. So all constraints and all combinations thereof just
represent the trivial conserved charge $Q\weak0$. Moreover, if we choose
different, but weakly equal representations for some conserved charge, then
they will generate different symmetry transformation, but the difference is
just a gauge transformation.

Quite generally, we can define a {\em reduced phase space\/} as the quotient of
the constraint surface $\phys$ with respect to gauge transformations. Then, an
observable simple becomes a function on the reduced phase space. There is also
a bracket on that space, as the Dirac bracket of two observables is again an
observable. If we go over to the reduced phase space, we can forget about all
the constraints and gauge transformations, it will just look like the phase
space of an unconstrained system. But in general it will not be a cotangent
bundle of some configuration space, and possibly it will not even be a proper
manifold, so it is much harder to deal with, especially when we try to quantize
it.

What are the observables and conserved charges in our examples? For the
particle on the circle, we found that the constraint surface was two
dimensional, so there are two independent observables and every observable is a
function thereof. They can be taken to be $\phi$ and $\pi=m\omega r$ as they
appear in \eref{c-con-sol}. There are many ways to express them in terms of
phase space function, for $\phi$ we can take $\arctan(-p_1/p_2)$ or
$\arctan(q_2/q_1)$ (or anything weakly equal). If we take $\pi^2=p_1^2 + p_2^2$
(some signs have to be fixed for the square root as well as for the $\arctan$,
but it should be clear how), then the {\em Poisson\/} bracket of the two
observables is obviously not well defined, as the value depends on which
representation we choose. But the {\em Dirac\/} bracket is, as you can check,
always $\{\phi,\pi\}_*\weak1$. The momentum $\pi$ is a conserved charge: the
generator of rotations.

For the two-dimensional gauge theory, the independent phase space functions
$p_1$, $p_2$ and $q_1+q_2$ have vanishing brackets with $\con$. However, $p_1$
and $p_2$ are weakly equal, so they represent the same observable and we are
left with two independent observables, $P\weak p_1\weak p_2$ and $Q\weak
q_1+q_2$. Their bracket is $\{Q,P\}\weak1$, so they are position and momentum
of a particle in one dimension. The reduced phase space is just that of a
particle in one dimension, and the Hamiltonian can be written as $\ham=\ft12
m^{-1} P^2 + \pot(Q)$.

\section{Quantization}

How do we quantize an unconstrained system? We usually take the configuration
variables and momenta as the basic observables and define a representation of
them as quantum operators on some Hilbert space $\hilb$. In quantum theory, a
state is represented by a vector in this {\em state space}. The operators have
to be chosen such that the commutator of two operators is given by the operator
corresponding to the Poisson bracket of the observables. What we need is a map
$\op{{\ }}$, that assigns an operator to each phase space function, such that
\beq[comm-pois]
  \comm {\op F}{\op G} =  \i \hbar \op{\pois FG}.
\eeq
And of course we want to have some other properties: the map should be linear,
the constant observables should be represented by scalar multiplication, and we
would also like to have something like $\op{{FG}}=\op F\op G$. This can in
general not be established completely. Simply because if $F$ and $G$ have
non-vanishing Poisson bracket, then the corresponding operators should not
commute, and therefore we cannot have $\op F \op G = \op{{FG}} = \op{{GF}} =
\op G \op F$. All we can do is to select some subset of observables, usually
taken to be a set of canonical variables like the $q$'s and $p$'s, and for them
we require \eref{comm-pois}. If we then give a general rule how to map phase
space functions onto operators, we find that \eref{comm-pois} will only hold
``modulo the operator ordering'' on the right hand side, which means that
\beq[comm-pois-h]
   \comm {\op F}{\op G} =  \i \hbar \op{\pois FG} + O(\hbar^2).
\eeq
There is always a {\em standard representation\/} if the phase space is the
cotangent space of some configuration space $\conf$. We can take the state
space to be the set of square integrable complex {\em wave functions\/} $\Psi$
on $Q$ and define
\beq[st-rep]
  \op F \, \Psi (\q) = F(\q) \Psi(\q), \qquad
  \op p_i \, \Psi(\q) = -\i\hbar \, \deldel \Psi / q^i / (\q),
\eeq
where $F$ is a function of $\q$ alone.
To get the operators for other functions, we have to expand them in a power
series in $\p$ to obtain the representation in terms of a power series in
$\del/\del\q$. So far this is standard quantum mechanics, but note then this is
just a special representation, and depending on the actual phase space given
there might be others which are more suitable.

There is one additional restriction we have to impose on the operator
representation, which ensures that our state space is not too big. The
representation should be irreducible, which means that the state space cannot
be decomposed into subspaces which are invariant under all observables. Then,
each subspace alone would already form a complete state space. Using the Dirac
notation $\kt\Psi$ for an element of the state space, and $\br \Psi$ for its
dual, we can briefly formulate standard quantum physics in the Schr\"odinger
picture as follows. Given a state $\kt{\Psi(t_0)}$, then it evolves in time as
\beq[psi-evol]
   \i \hbar \, \dd / t/ \ket {\Psi(t)} = \op \ham \, \ket{\Psi(t)}.
\eeq
Statements about measurements and physical properties of a state are made in
form of {\em expectation values\/} of observables
\beq[exp-val]
  \expect{F} = \bra \Psi \op F \ket \Psi.
\eeq
As we want expectation values corresponding to real phase space functions to be
real, the scalar product has to be such that real observables are represented
by Hermitian operators. Note that this is not a restriction on the operator
representation $\op{{\ }}$ or on the scalar product on $\hilb$ alone, but it
states a {\em relation\/} between them. This will be quite important for
constrained systems. In the standard representation \eref{st-rep}, the unique
scalar product is
\beq[st-sp]
  \braket \Phi \Psi = \int \! \d\q \ \ \Phi^*(\q) \Psi(\q),
\eeq
where $\d\q$ is the measure that is invariant under ``translations'' generated
by the $\op p_i$ operators. Of course, this only exists if the configuration
space in finite dimensional. If this is not the case, we have to use a
different definition of the scalar product, as we shall see in our examples in
the next section.

We also have theorems like Ehrenfest's, which essentially tells us that
expectation values behave {\em almost\/} like the classical phase space
function. For example, we have
\beq
  \expect{\dt F} = \bra \Psi \op{\pois F\ham} \ket \Psi =
   \frac 1{\i\hbar} \bra \Psi \comm{\op F}{\op\ham} \ket \Psi + O(\hbar) =
          \deldel / t / \expect{F}  + O(\hbar).
\eeq
We can define all this for a constrained system as well. We can even take the
standard representation \eref{st-rep} and what we get is a state space, let us
call it $\stsp$, and a representation of the phase space functions as linear
operators. We also get a time evolution of the state, generated by the
Hamiltonian. If there are gauge degrees of freedom, it will contain free
parameters, and they will also appear in the quantum theory. The time evolution
will not be unique. So far this seems to be quite straightforward. But let us
have a look at Ehrenfest's theorem. If we want that the quantum expectation
values behave like the classical phase space functions, this should especially
hold for the constraints. So, we would like to have
\beq[exp-con]
  \bra \Psi \op\con_\alpha \ket \Psi = 0.
\eeq
We can take these quantum constraints as a restriction to be imposed on the
state $\kt \Psi$. In analogy to the classical counterparts we can call them
{\em physical states\/}. The problem with this restriction is that it does not
single out a linear subspace of $\stsp$, as it is a quadratic equation in
$\Psi$. Moreover, what about expectation values of the form
\beq[exp-con-2]
  \bra \Psi \op\con_\alpha \op\con_\beta \ket \Psi \, ?
\eeq
If we want Ehrenfest's theorem to hold, these have to vanish as well, and the
same must hold for every function of the constraints that classically vanish,
at least to the leading order in $\hbar$. If all these expectation values
vanish, then we have an eigenstate, and the condition to be imposed on a
physical state becomes
\beq[q-phys]
  \op\con_\alpha \ket \Psi = 0.
\eeq
This provides a linear subspace, which is called the {\em physical state
space}. At first sight, it is not clear that an equation like this can actually
be solved, because it states an eigenvalue equation for $\kt\Psi$. Perhaps the
operator in front doesn't have zero as an eigenvalue. If this is the case,
there is no other way out than to choose a different operator representation.
However, things are not as restrictive as they might seem. The representation
space $\stsp$ is not required to be a Hilbert space: there is no ``finite
norm'' restriction on $\kt\Psi$, nor is the constraint operator Hermitian or
whatsoever. Using the standard representation, \eref{q-phys} becomes a set of
differential equations, but without any boundary conditions, which is in
general easy to solve.

If we take our simple gauge system as an example, the standard representation
is given by $\Psi(q_1,q_2)$, and $\op p_i=-\i\hbar\del/\del q_i$ for $i=1,2$.
The quantum constraint is
\beq[q-ex-con]
  \deldel \Psi / q_1  /  = \deldel \Psi / q_2 / .
\eeq
Clearly, the general solution is $\Psi=\Psi(q_1+q_2)$. This shows two crucial
properties of constrained systems. The first is that such a function can never
be square integrable on $\RR^2$. Generally, for constrained systems the
standard product will not give a suitable scalar product on the physical
states. A second property of the wave function its gauge invariance. Remember
that in the classical theory we had to perform two steps to get a
gauge-invariant object: we had to solve the constraints, and then we had to
consider equivalence classes of states generated by the first class
constraints. Here, it seems that the first step already leads to a gauge
invariant object. This also holds quite generally, but let us first consider
the conditions \eref{q-phys} in more detail.

There might be another problem when we try to solve them. It is not obvious
that the equations are compatible. They imply
\beq[com-con]
  \comm{\op\con_\alpha}{\op\con_\beta} \ket \Psi = 0,
\eeq
and this should not impose any new restriction on the state. For the classical
constraints we know that their Poisson bracket is again a linear combinations
of constraints, if at least one of them is a first class constraint. But what
happens if we take two second class constraints in \eref{com-con}? In the worst
case the commutator will be the identity operator (as in the trivial example
$\scl_1=q^1$, $\scl_2=p_1$), and the condition gives $\kt\psi=0$. But it should
now be clear how to proceed. At the classical level, we solved the problems
concerning second class constraints by introducing the Dirac bracket. We can do
the same here. Instead of defining the operators to be a representation of the
Poisson bracket algebra, we use the Dirac brackets everywhere. This should work
equally well, as they have the same basic properties as the Poisson brackets.
Except that now we cannot choose the standard representation any more.

Hence, if we have second class constraints, they have to be implemented in the
quantum theory from the very beginning. We can impose them as exact operator
identities, as the Dirac brackets were made such that the second class
constraints behave like identities. The physical state condition is then
trivially satisfied for them, we only have to impose it for the first class
constraints. For the particle on the circle this means that we can choose the
state space to be the set of complex wave functions on the circle. The
operators are then given by \eref{c-con-sol}, with $\omega$ replaced by a
suitable differential operator with respect to $\varphi$. This is of course
where we also would have ended up if we had started form the simpler Lagrangian
depending on $\varphi$ only, describing a free particle on a circle. So, as at
the classical level, the second class constraints are not of much interest. In
a sense, they result from choosing unsuitable variables. In the following we
will therefore assume that there are no second class constraints present.
Otherwise just replace the Poisson brackets by Dirac brackets everywhere.

If all constraints are first class, then we can expect \eref{com-con} to be a
linear combination of the equations \eref{q-phys}, provided that the classical
algebra of the first class constraints \eref{gauge-alg} is preserved:
\beq[q-gauge-alg]
  \comm{\op\fcl_a}{\op\fcl_b} = \i\hbar\, \op f_{ab}\^c \, \op\fcl_c.
\eeq
The structure functions $f_{ab}\^c$ in the classical algebra are in general
phase space functions, so in the quantum theory they are represented by
operators. These do not necessarily commute with the constraints. So it is
essential that they appear to the left. Otherwise \eref{com-con} gives an
equation for physical states that has no classical counterpart, and it might
even give a contradiction as we saw for the second class constraints. The
``structure operators'' $\op f_{ab}\^c$ need, however, not be the operators
associated with the classical structure functions, it is only necessary that
they exist and appear to the left. If it is not possible to give an operator
representation such that \eref{q-gauge-alg} holds, we cannot impose
\eref{q-phys}, or if we impose it we would get a too small physical state
space. This is a non-trivial problem in Dirac's canonical quantization. If the
constraints can be arranged in this way, then \eref{q-gauge-alg} gives the
quantum representation of the gauge algebra. In typical gauge field theories,
where the structure functions are numbers and the gauge transformations act
linear on the fields, this can always be achieved. If the gauge group is
however more involved, e.g.\ the diffeomorphism group in gravity or string
theory, it is not clear that we can always find a representation satisfying
\eref{q-gauge-alg}. There may however be a way out as we shall see in a moment.

Let us first consider the time evolution. Using the decomposition
\eref{ext-ham} for the Hamiltonian, we have
\beq[q-evol]
   \i \hbar  \, \dd / t /  \ket \Psi =
   \op \phm \, \ket \Psi + u^a  \, \op \con_a \, \ket \Psi.
\eeq
Actually the $u^a$ should become operators as well, as we were free to insert
any phase space function for them in the classical theory. However, for
physical states $\kt\Psi$ the last term vanishes anyway, and we are left with
the first one only. To get a consistent theory, this must not take us away from
the physical states. As the state space is a linear space, we can simply check
this by acting on the time derivative of $\kt\Psi$ with a constraint and see
whether the result vanishes.
\beq
  \op \fcl_a \, \op \phm  \ket \Psi  =
  \comm{\op \fcl_a }{\op \phm } \ket \Psi  -
  \op \phm \, \op \fcl_a  \ket \Psi  .
\eeq
Clearly, the last term vanishes as $\kt\Psi$ is a physical state. At the
classical level, we know that $\ham$ is first class, so its bracket with any
constraint vanishes weakly. Here, we get an extra condition, namely that this
property is preserved in the quantum theory:
\beq[q-ham-fcl]
  \comm{\op\fcl_a}{\op\phm} = \i\hbar\, \op g_a\^b \, \op\fcl_b,
\eeq
with the coefficients $\op g_a\^b$ again appearing to the left.

There is something else we can learn from \eref{q-evol}, and we saw this
already for our simple example. In the classical theory, the last term was the
one that generated gauge transformations on the state. Here we find that it
simply vanishes when it acts on a physical state. Physical states are already
gauge invariant quantities. Imposing \eref{q-phys} somehow combines the two
steps we needed at the classical level. There, we had to impose the constraints
and then identify those states that are related by a gauge transformation. In a
sense, these two operations are conjugate to each other. To illustrate this,
let us try to interchange them. At the classical level, we could equally well
{\em first\/} take equivalence classes and then restrict to the constraint
surface. This would just result in defining some classes outside the constraint
surface as well, which will afterwards be thrown away anyway.

In quantum theory, starting from the representation space $\stsp$, we can
consider two states as gauge equivalent if they can be deformed into each other
by a transformation generated by the constraints. As the states live in a
vector space, we can simply declare two states to be equivalent if their
difference lies in the image of the constraint operators:
\beq[q-ekl]
  \ket{\Psi} \sim \ket{\Phi} \equiv
  \ket{\Psi} - \ket{\Phi} =  \op\fcl_a \ket{\Omega^a}
\eeq
for some vectors $\kt{\Omega^a}$. Then, we can define the physical state space
as the quotient space of $\stsp$ with respect to this equivalence relation.
Intuitively, it should be clear that this will result in the same physical
state space as the other procedure. Building the equivalence classes is
effectively the same as imposing the constraints on the {\em dual\/} vector
space\footnote{A dual vector can be considered as a projector onto some
one-dimensional subspace of the state space, and if we consider only those
projections satisfying \eref{c-dual-cond}, then the states in \eref{q-ekl}
effectively become equal because there is no projection that can distinguish
them.}
\beq[c-dual-cond]
    \bra\Psi \, \op\fcl_\alpha  = 0.
\eeq
In a sense, we are interchanging the kernel with the image of the constraint
operators. The problems occurring are the same in both cases. If we cannot find
a representation where the first class constraints form a closed algebra, the
equivalence classes will become too big and therefore the physical state space
too small. The only difference is that now we have to require the ``dual'' of
\eref{q-gauge-alg}, i.e.\ the structure functions have to appear {\em to the
right\/} as
\beq
  \comm{\op\fcl_a}{\op\fcl_b} = \i\hbar\, \op\fcl_c \, \op f_{ab}\^c .
\eeq
It seems that the ``operator ordering problem'', i.e.\ the problem of finding a
representation that preserves the classical constraint algebra, cannot be
solved in this way either. We just have to reverse all the operator ordering to
get from one picture to the other. However, in many cases where this problem
actually occurs, there is a way to solve it by combining the two procedures.
Assume that it is possible to split the complete set of (first class)
constraints into two subsets, denoted by $\fcl_a$ and $\fcl\cc_a$. In many
cases\footnote{An important example where this trick works is string theory,
where the constraints form a Virasoro algebra, whose central charge violates
\eref{q-gauge-alg}, but \eref{q-cc} holds after splitting them into
``positive'' and ``negative'' frequency operators.} these turn out to be
complex conjugate sets of constraints, but as long as we do not have a Hilbert
space this is of no significance here. If they can be chosen such that
\beq[q-cc]
 \comm{\op\fcl_a}{\op\fcl_b} = \i\hbar\, \op f_{ab}\^c \, \op\fcl_c ,\qquad
 \comm{\op\fcl{}\cc_a}{\op\fcl{}\cc_b}
       = \i\hbar\, \op\fcl{}\cc_c \, \op f\cc_{ab}\^c ,
\eeq
then we can apply a mixture of the two methods. First, we impose \eref{q-phys},
but only for one half of the constraints, which is consistent by the first
equation. Then, we take equivalence classes with respect to the second set.
These equivalence classes will in general not be ``tangent'' to the result of
the first step. In other words, adding something of the form
$\op\fcl{}\cc_a\kt{\Omega^a}$ to a physical state may result in an unphysical
one. But that does not cause any problems, it simply means that the equivalence
classes are smaller. Here is a simple example. Consider two first class
constraints $\fcl$ and $\fcl^*$, such that their classical Poisson bracket
vanishes but $\comm{\op\fcl}{\op\fcl{}^*}\propto\hbar^2$. Clearly,
\eref{q-gauge-alg} does not hold, but \eref{q-cc}. Solving $\op\fcl\,\kt\Psi=0$
gives the physical states. Now, declare two such states as equivalent, if their
difference is $\op\fcl{}^*\kt\Omega$ for some physical state $\kt\Omega$. You
can easily see that this never happens except for $\kt\Omega=0$. The
equivalence classes are trivial and we get a well defined physical state space.

If we use this procedure, we do not have the strong condition \eref{q-phys} any
more, but Ehrenfest's theorem still holds for the constraints. In
\eref{exp-con}, the constraints will either annihilate the state to the left or
to the right. For the higher order terms \eref{exp-con-2}, we probably have to
``normal order'' the constraints to make them act on the correct side, but this
produces corrections of order $O(\hbar)$ only, and these cannot be avoided
anyway in Ehrenfest's theorem. So, even if we cannot define a proper algebra of
quantum constraints, there are ways to overcome this problem. For simplicity
however, let us in the following assume that we succeeded in finding a
representation such that \eref{q-gauge-alg} holds, and that there is a well
defined physical phase space, denoted by $\hilb$.

It is not a Hilbert space yet, so we have to define a scalar product. We want
that the real observables are represented as Hermitian operators. In the
classical theory, observables were found to be those phase space functions
whose brackets with the constraints vanish weakly. Let us introduce a similar
``weak equality'' in quantum physics. We call two operators {\em weakly
equal\/} if there difference is a linear combination of constraints, with the
coefficients appearing to the left,\footnote{If we use the ``mixed scheme''
introduced above, the $*$-constraints must appear to the left and the others to
the right, and similar modification are to be made in what follows.}
\beq
  \op F \weak \op G  \equiv
   \op F - \op G = \op u^a \op \fcl_a.
\eeq
Remember that the second class constraints have been realized as exact operator
equalities, so they don't show up here. As an example, the condition
\eref{q-ham-fcl} can now be expressed as
\beq
  \comm{\op\phm}{\op\fcl_a} \weak 0.
\eeq
A quantum observable $\op O$ is defined to be an operator with exactly this
property: it has to commute weakly with all constraints
\beq[q-obs-def]
  \comm{\op O}{\op\fcl_a} \weak 0.
\eeq
It is not immediately obvious the every classical observable can be related to
a quantum observable, again because of the ordering problems which are present
every time we are replacing a bracket by a commutator. But let us assume that
we can find a representation such that the operator for any classical
observable becomes a quantum observable. The fact that the commutators of
observables with  constraints vanish weakly tells us that the observable maps
physical states onto physical states, as
\beq
  \op\fcl_a \, \op O   \ket \Psi =
    \op O \,  \op\fcl_a \ket \Psi
  - \comm{ \op O }{\op\fcl_a}  \ket \Psi = 0.
\eeq
Here we used that every weakly vanishing operator, acting on a physical state,
gives zero. The observables are exactly those operators that act on the
physical state space $\hilb$. When restricting the operators on $\hilb$, the
weak equality becomes the identity. As for the classical theory, observables
are actually equivalence classes of operators modulo the weak equality. If we
want all this to be in analogy with the unconstrained system, $\hilb$ should
become the physical Hilbert space, so that we can compute expectation values of
observables using \eref{exp-val}. As only observables represent physical
quantities, there is no need to have expectation values for any other operator,
and it is therefore sufficient to have a scalar product only on the physical
state space. We saw already that sometimes it is even {\em impossible\/} to
extend the product to the original representation space $\stsp$.

This final step of the Dirac programme turns out to be a rather sophisticated.
There is no general recipe how to do it. In most cases it is pretty obvious how
the product has to look, but sometimes it is not at all clear. All we have to
fix it is that we require the real observables to become Hermitian operators,
or equivalently that the complex conjugate of a classical observable is
represented by the Hermitian conjugate of the quantum observable. This will
ensure that Ehrenfest's theorem also holds for the conjugation:
\beq
  \expect{F}^* = \expect{F^*} \equiv
  \op F^\dagger = \op{F^*}.
\eeq
As for unconstrained systems, the last equation gives a relation between the
representation map $\op{{\ }}$ and the scalar product which defines the
conjugation map $^\dagger$. But is also imposes some extra restrictions on
$\op{{\ }}$ alone. If we are unlucky, all we did so far was in vain, because
these are not realized. Suppose, for example, that there are three real
classical observables satisfying $AB=C$, and the operator representation gives
$\op A\op B=\op C$. If $\op A$ does not commute with $\op B$, this can never be
a relation between Hermitian operators. So, when applying Dirac's quantization
scheme, it is a good idea always to keep in mind this problem, which might
occur in the last step. We should take care in all intermediate steps and make
sure that they are compatible with the $*$-relation.

{}From this point of view, the ``mixed'' scheme for the physical state
conditions discussed above becomes more ``natural'' as well. To see why we
should use two complex conjugate set of constraints in \eref{q-cc}, consider
the following. When we are talking about complex constraints there is a little
subtlety. At the classical level, we always considered real equations of the
from $\fcl_a(\q,\p)=0$. Now assume that we can combine the constraints into a
set of half as many complex equations. We can do this whenever their number is
even, but let us assume that there is a ``reasonable'' way to do so. Then the
classical conditions are still the same as before, only their form has changed:
they are half as many but complex. We can write them equivalently as $\fcl_a=0$
or $\fcl\cc_a=0$. But this is no longer true at the quantum level, if we impose
the conditions \eref{q-phys}. Should we impose $\op\fcl_a\kt\Psi=0$ or
$\op\fcl{}\cc_a\kt\Psi=0$, or both? It won't have any impact on expectation
values, up to orders in $\hbar$. Indeed, imposing only the first condition is
just what we did when we used the mixed framework. That is because, if we have
a scalar product, then $\op\fcl_a\kt\Psi=0$ becomes equivalent to
$\br\Psi\op{\fcl\cc}_a=0$. What the mixed scheme actually does is to impose
only half of the constraints. This turns out to be enough, because the physical
state condition $\op\fcl_a\ket\Psi=0$ is always a complex equation. It is the
quantum counterpart of the complex classical equation $\fcl_a(\q,\p)=0$. There
is no need to impose real and imaginary part separately.

In most cases, it is either possible to impose all constraints as
$\op\fcl_a\kt\Psi=0$, or they arrange properly in two sets of complex conjugate
constraints, so that one of the scheme can always be applied. However,
sometimes this final step really causes serious problems, and this seems to be
the case in some approaches to quantum gravity. The maybe famous one in the
moment is the Ashtekar programme. It turned out that by choosing suitable
variables, the constraints to be solved become rather simple, and many physical
states could be found. But the variables are such that during the quantization
procedure one gets completely lost of the complex conjugation relations. The
conjugation operation is realized non-analytically, which leads to the problem
that if $\op F$ can be given as a well quantum operator, $F^*$ will in general
not have a proper operator representation, as it cannot be expanded in a power
series of differential operators. It is clear that this makes it hard to
construct the correct scalar product, and it is this point where the Ashtekar
programme got stuck.  So the most non-trivial step in Dirac's quantization
programme is the last one: to find the scalar product on the physical state
space.

\section{Examples}
I will consider three examples here, all having gauge degrees of freedom, but
of very different types. The first is the free electro-magnetic field. Here,
the gauge group is rather simple, but we have to deal with a field theory which
has an infinite dimensional phase space. This example will show that Dirac's
programme will give us a manifestly gauge covariant quantization, in contrast
to most of the ``modern'' methods, where a gauge fixing is necessary before it
can be quantized.

The second example is the relativistic particle in a background
electro-magnetic field. It is the simples example for a system with {\em
parameter time}. This means that the time coordinate that appears in the
integral that defines the action is not the physical time, but rather some
parameter that can be chosen arbitrarily. The possible reparametrizations of
that coordinate will show up as gauge symmetries in the Dirac programme.

Finally, in string theory, considered as a two dimensional field theory, we
also have a parameter time, but in addition we have infinitely many degrees of
freedom. It turns out that the resulting constraints can no longer be quantized
such that their classical algebra is preserved, and as a consequence we have to
use the ``mixed scheme'' to define the physical state space.

\subsection*{The electro-magnetic field}
This is the simplest example for a non-trivial gauge field theory. Its action
is given by
\beq[em-act]
\act[A] = - \ft14 \intd4x  F^{\mu\nu} F_{\mu\nu}, \qquad
          F_{\mu\nu} = \del_\mu A_\nu - \del_\nu A_\mu.
\eeq
The spacetime metric is taken to be $g^{\mu\nu}=\diag(-1,1,1,1)$. We have to
write this as an integral of some Lagrangian over the time. To do this, we have
to choose a time coordinate, which we shall take to be the $0$ coordinate of
Minkowski space. The index $\mu$ will be split into the $0$ component and the
remaining spatial components $i=1,2,3$. Similarly, the potential $A_\mu$ splits
into a time component $A_0$ and the space components $A_i$. The spatial indices
can always be written as lower indices, as the metric for them is just the unit
matrix. The Lagrangian becomes
\beq[em-lag]
  \lag[A,V] =\intd3x  \ft12 F_{0i} F_{0i} -  \ft14 F_{ij} F_{ij},
   \qquad
    F_{0i} = V_i - \del_i A_0, \quad
    F_{ij} = \del_i A_j - \del_j A_i.
\eeq
Here, $V_\mu$ is the ``velocity'' $\del_0 A_\mu$. The configuration space is
given by the set of {\em spatial field configurations\/} $A_\mu(x)$, and a time
evolution is given by a {\em spacetime field configuration\/} $A_\mu(x,t)$,
where $t=x^0$ is the coordinate over which the Lagrangian has to be integrated
to give the action. This shows a general problem of the Dirac programme when
applied to field theories. One has to single out a special time coordinate,
thereby breaking the manifest Lorentz covariance of expressions like
\eref{em-act}. At the classical level, there exists a generalization of the
Dirac programme, the {\em De~Donder Weyl\/} canonical formalism, which does not
break this symmetry and treats all spacetime coordinates similarly. Instead of
a spatial field configuration evolving in time, it in some sense considers a
field ``evolving'' in spacetime. The problem with this nice formalism is that
up to now nobody has been able to construct a quantum theory based thereon. So,
here we have to stick to this ``non-relativistic'' formalism, where some
coordinate $t=x^0$ is distinguished.

The momenta are found to be
\beq[em-mom]
  E_i(x) = \deltadelta \lag / V_i(x) / = F_{0i}(x), \qquad
  E^0(x) = - E_0(x) = \deltadelta \lag / V_0(x) / = 0.
\eeq
This gives a primary constraint
\beq[em-prim]
  \fcl_1(x) = - E_0(x).
\eeq
Note that this is not just a single constraint, but a infinite set of
constraints, one for each space point. The point $x$ now plays the role of the
index $\alpha$ in \eref{prim-ham}. This holds for the momenta as well: there is
a momentum $E^\mu(x)$ conjugate the field $A_\mu(x)$ at each space point. For
the Poisson bracket we have
\beq[em-pois]
   \pois{ A_\mu(x) } {E_\nu(y)} = g_{\mu\nu} \delta^3(x,y).
\eeq
The Hamiltonian is easily found to be
\beq[em-phm]
  \ham = \intd3x  \ft12 E_i E_i + \ft14 F_{ij}F_{ij}
                 + E_i \, \del_i A_0  + u \, E_0,
\eeq
with one free parameter $u(x)$ for each $x$. The time evolution equations are
\beq[em-evol]
   &&\dt A_i = \pois{A_i}{\ham} = E_i + \del_i A_0, \qquad
   \dt A_0 = \pois{A_0}{\ham} = - u, \zl
   &&\dt E_i = \pois{E_i}{\ham} = - \del_j F_{ij}, \qquad
   \dt E_0 = \pois{E_0}{\ham} = - \del_i E_i .
\eeq
Here it should be clear how the arguments $x$ are to be added, so I didn't
write them explicitly. The last equation tells us that there is a secondary
constraint
\beq[em-sec]
  \fcl_2(x) = - \del_i E_i(x).
\eeq
As you can read off from the definition of the momenta, and also from the
evolution equations, $E_i$ is of course nothing but the {\em electric field}.
There is one equation for $E_i$ which doesn't contain time derivatives, namely
the Gau\ss\ law, which has to show up as a constraint. There are no more
constraints, as
\beq
  \pois{\con_2}{\ham} = -\del_i \del_j F_{ij} = 0
\eeq
by antisymmetry. Especially, there are no restrictions on $u$. It is also
immediately obvious that the constraints are first class, as they only depend
on the electric field and therefore their bracket vanishes. Hence, we expect
them to generate gauge transformations. Consider a general ``linear
combination'' of constraints
\beq
  \fcl[\u] = \intd3x u_1(x) \fcl_1(x) + u_2(x) \fcl_2(x),
\eeq
where $\u$ is a pair of scalar functions on space. Then, we have
\beq[em-gauge]
  \delta A_0 = \pois{A_0}{\fcl[\u]} = u_1 , \qquad
  \delta A_i = \pois{A_i}{\fcl[\u]} = \del_i u_2,
\eeq
and the electric field in invariant.
Clearly, this is just what we know to be a gauge transformation in
electrodynamics. However, there is one point which should be mentioned here. If
we look at the theory from the four dimensional point of view, a gauge
transformation should be given as $\delta A_\mu=\del_\mu u$. So, actually the
two parameters in\eref{em-gauge} should be related by $u_1=\dt u_2$. From the
point of view we are taking here, these two are independent as we are only
considering gauge transformations at a special moment of time. You can check
that this is in agreement with our definition of gauge transformations. Using
the time evolution \eref{em-evol}, you can achieve any transformation generated
by \eref{em-gauge} by going a small amount backwards in time, and then forward
again with a differently chosen $u$.

Here we have what we already discussed quite generally. In the Hamiltonian,
only the primary first class constraints appear with free parameters. But the
secondary ones also generate gauge transformations and so we are free to add
them to the Hamiltonian, with free parameters in front. This gives the total
Hamiltonian
\beq
  \ham = \intd3x  \ft12 E_i E_i + \ft14 F_{ij}F_{ij}
                 + u_1 \, E_0  - u_2 \, \del_i E_i ,
\eeq
where the $E\del A$ term has been absorbed by redefining $u_2\mapsto u_2+A_0$.
As we concluded quite generally, this extended Hamiltonian no longer generates
solutions to the original equations of motion (except if we choose $u_2$ to be
$A_0$). But physically we cannot distinguish them because they will always be
gauge-equivalent to some extremum of the action.

What are the observables? We already found that the electric field is an
observable, as $\{E_i,\fcl[\u]\}=0$. $E_0\weak0$ is also an observable, but it
is a rather trivial one. The remaining observables are the {\em magnetic
fields\/} $B_i=\ft12\eps_{ijk}F_{jk}$. There is no observable that depends on
$A_0$, because it can be completely ``gauged away'' by \eref{em-gauge}. The
energy is given as the value of $\ham$ on the constraint surface, which is
\beq
  \ham \weak \intd3x \ft12 E_i E_i + \ft12 B_i B_i.
\eeq
To quantize the field, let us choose the standard representation. The state
$\kt\Psi$ is thereby given as a {\em wave functional\/} $\Psi[A]$ of the gauge
potential. Clearly, $\op A_\mu$ acts as a multiplication operator and the
electric field as
\beq[qed-mom]
  \op E_\mu(x) \, \Psi = - \i \hbar g_{\mu\nu} \, \deltadelta \Psi /A_\nu(x)/.
\eeq
Imposing the constraints means that we have to require
\beq[qed-con]
  \op \fcl_1 \ket \Psi = 0 \equiv \deltadelta \Psi / A_0(x) / = 0 , \qquad
  \op \fcl_2 \ket \Psi = 0 \equiv \del_i \deltadelta \Psi / A_i(x) / = 0 .
\eeq
The first one is trivially solved: $\Psi$ must not depend on $A_0$. The second
states that the state has to be a gauge-invariant functional. This is a general
feature of standard gauge theories. The constraints are linear functions of the
momenta, and if we choose the standard representation, the action of the
constraints on the wave functional becomes a simply gauge transformation. Note,
however, that this holds only as long as the constraints are linear in the
differential operators. So, effectively the constraints tell us that a physical
state must be a gauge-invariant functional.

We can now apply another rather general technique to get an overview over the
physical state space. It works as follows. Assume that we found some special
physical state by explicitly solving the constraint equations. Then, we can act
on that state with an observable, which will again give a physical state. As
the physical state space is required to be an irreducible representation of the
observable algebra, we can generate every physical state in this way. The
result will be some kind of Fock space. Let us try it. Here is a special
gauge-invariant wave functional, the Chern Simons form
\beq
   \Psi_0[A] =
      \exp \Big( - \frac 1{2\hbar} \intd3x \eps_{ijk} \, A_i \del_j A_k \Big).
\eeq
Let us denote the corresponding state by $\kt0$. It is, up to normalization,
the unique solution to the functional differential equation
\beq[vac-q]
   \big( \op E_i - \i \op B_i \big) \ket 0 = 0 .
\eeq
The operator here is an observable, and together with its conjugate $\op E_i +
\i \op B_i$ we have a complete set of observables. They are complex conjugate,
so they might be good candidates to generate a Fock space. Choosing a suitable
operator ordering, the Hamiltonian can also be written in a nice form as
\beq
  \op \ham = \intd3x \ft12 ( \op E_i + \i \op B_i )( \op E_i - \i \op B_i ).
\eeq
This is exactly what we need for the typical Fock representation, as now we
have $\op\ham\kt0=0$. So is $\kt0$ the vacuum state? Everything looks pretty
nice up to now, but yet we haven't done the last step. The scalar product on
the state space is still missing. How shall we define it? We could try
\beq
  \braket \Phi\Psi = \int\!\d A \ \ \Phi^*[A] \Psi[A],
\eeq
but a functional integral like this is really hard to deal with. In particular,
we would get all the problems present in the path integral approach. Moreover,
this product would, if it exists, be defined on the whole state space, not just
on the physical state space, and we saw that this is in general impossible.
However, we have something like a Fock space, so we should try to define it
recursively. We normalize it by
\beq
  \braket 0 0 = 1.
\eeq
Now consider the ``one photon'' state
\beq
  \ket \Omega=
    \intd3x  \Omega_i(x) \big( \op E_i(x) + \i \op B_i(x) \big) \ket 0,
\eeq
which is physical,because the operator is an observable (its wave functional is
$\int2\i\Omega^iB_i\,\Psi_0$ and solves \eref{qed-con}). We want our scalar
product to preserve the complex conjugation, which means that the Hermitian
conjugate relation
\beq
  \big( \op E_i(x) + \i \op B_i(x) \big)^\dagger =
  \big( \op E_i(x) - \i \op B_i(x) \big)
\eeq
should hold. Using this we can compute the norm of the new state
\beq
  \braket \Omega\Omega= \int \!\d^3x\,\d^3y \,\,\Omega\cc_i(x) \Omega_j(y)\,
  \bra 0 \big( \op E_i(x) -  \i \op B_i(x) \big)
      \, \big( \op E_j(y) +  \i \op B_j(y) \big) \ket 0.
\eeq
Using the commutator
\beq
   \comm{E_i(x)}{B_j(y)} = - \i \hbar\,\eps_{ijk} \,\del_k \delta^3(x-y),
\eeq
and \eref{vac-q}, this gives
\beq[sp-fail]
   \braket \Omega\Omega= 2\hbar \intd3x \eps_{ijk}
    \Omega\cc_i(x) \del_j \Omega_k(x).
\eeq
This is not positive definite. By choosing $\Omega_i$ suitably, it can become
any negative number. What does this mean? Well, it tells us that it is not
possible to define a scalar product such that the state $\kt0$ is normalizable
{\em and\/} the complex conjugation relations for the observables are correctly
realized as Hermitian conjugation relations. These were the two assumptions we
made. To get a correct quantum theory, we have to find a different ``vacuum
state'', i.e.\ another solution to \eref{qed-con}, to build a Fock space on.
One needs to play a bit around to find the correct one that reproduces the
standard Fock space of quantum electro-dynamics. It is
\beq[qed-vac]
   \Psi_0[A] = \exp \Big( -\frac 1{2\hbar} \intdkk B_i(k) B_i(-k) \Big),
\eeq
where $B_i(k)$ is the Fourier transform
\beq[f-trans]
   B_i(k) = \intd3x e^{-\i k\cdot x} \, B_i(x) = \i \eps_{ijk} \, k_j A_k(k),
\eeq
and similar definitions will be used for all other fields. The commutators of
the transformed operators are
\beq[f-com]
  &&\comm {\op E_i(k)}{\op A_i(l)} =
   - \i \hbar (2\pi)^3 \, \delta_{ij} \, \delta^3(k+l), \zl
  &&\comm {\op E_i(k)}{\op B_j(l)} =
   - \hbar (2\pi)^3 \,  \eps_{ijk} k_k \, \delta^3(k+l) .
\eeq
This give the following operator for the electric field when it acts on a
functional of $A_i(k)$:
\beq
  \op E_i(k)\Psi = - \i \hbar (2\pi)^3 \, \deltadelta \Psi / A_i(-k) / .
\eeq
The Hamiltonian is also straightforwardly transformed as
\beq
  \ham & =& \intdkkk E_i(-k) E_i(k) +  B_i(-k) B_i(k) \zl
        & =& \intdkkk E\cc_i(k) E_i(k) +   B\cc_i(k) B_i(k) .
\eeq
To find the creation an annihilation operators, we act with $\op E_i(k)$ on the
vacuum state \eref{qed-vac}, which gives
\beq[qed-vac-eq]
  \Big(\op E_i(k) +  \eps_{ijk}  \frac{k_j}{|k|} \op B_k(k) \Big)
   \ket 0 = 0 .
\eeq
So this is expected to be the annihilation operator. We define
\beq[qed-a-a]
  \op a_i (k) =     \op E_i(k) + \eps_{ijk} \frac{k_j}{|k|} \op B_k(k), \qquad
  \op a\dgg_i(k) = \op E_i(-k) + \eps_{ijk} \frac{k_j}{|k|} \op B_k(-k).
\eeq
They also form a complete set of independent observables as we can solve these
equations for $E_i$ and $B_k$ (except for $k=0$, but $B_k(0)=0$ anyway). They
are conjugate to each other and the only non-vanishing commutator gives
\beq[qed-a-com]
  \comm {\op a_i(k)}{\op a\dgg_j(l)} =
    2 \hbar (2\pi)^3 \, \frac{\delta_{ij} k^2 - k_ik_j}{|k|} \, \delta^3(k-l).
\eeq
The crucial property of this commutator is that the matrix appearing on the
right hand side is positive semidefinite. It is now possible to define the
scalar product. We repeat the construction from above with
\beq[qed-cr]
  \ket \Omega= \intdk  \Omega_i(k) \, a\dgg_i(k) \ket 0,
\eeq
Using the commutator \eref{qed-a-com} the norm of this vector becomes
\beq
   \braket \Omega\Omega= 2\hbar \intdkk
    \Omega\cc_i(k) \Omega_j(k) \big(k^2 \delta_{ij} - k_i k_j \big).
\eeq
In contrast to \eref{sp-fail}, this will never be negative. It follows that for
all physical states generated by the $\op a\dgg_i$ operators, we have
$\brkt\Psi\Psi\ge0$. What remains to be shown is that the only state with zero
norm is $\kt\Psi=0$. The integral above becomes zero if an only if $\Omega_i(k)
=  k_i\Omega(k)$. Inserting this into \eref{qed-cr} we get
\beq
  \ket \Omega=
    \intd3x  \Omega_i(k) \, \op a\dgg_i(k) \ket 0 =
    \intd3x   \Omega(k) \, k_i\, \op E_i(k) \ket 0 = 0,
\eeq
because $k_i E_i(k)$ is nothing but the Fourier transform of the constraint
$\fcl_2(k)=\i k_i E_i(k)$. The same holds if we replace $\kt0$ by any physical
state, and therefore our scalar product is perfectly well defined. We also see
that the function $\Omega_i$ appearing in the creation operator \eref{qed-cr}
has some ``gauge freedom''. It creates the same physical state if we replace
$\Omega_i(k)\mapsto\Omega_i(k)+k_i\lambda(k)$. So we might gauge fix by
requiring $k_i\Omega_i(k)=0$, which means that there should be no
``longitudinal modes'' in $\Omega$. However, this is only necessary if we want
to {\em classify\/} the physical states, say, by defining a Fock basis of the
physical state space. Such a gauge fixing is completely unnecessary for the
{\em definition\/} of the state space. The creation operators take perfectly
care that we never create an unphysical ``longitudinal'' state, simply because
they are observables. The gauge freedom just shows up as some ambiguity in
$\Omega$: choosing different functions changes the operator only {\em weakly}.

Finally, we should also check whether the $\op a^\dagger_i$ operators really
create the correct eigenstates of the Hamiltonian, with positive eigenvalues.
We can express $\ham$ in terms of the creation and annihilation operators. We
choose the correct operator ordering to get
\beq[qed-ham]
  \op \ham = \intdkkk  \op a\dgg_i(k) \, \op a_i(k) \follows
   \op \ham \ket 0 =  0.
\eeq
Now we can compute the commutator of a creation operator with the Hamiltonian
\beq
  \comm{\op\ham}{\op a\dgg_i(k)}
   = \hbar\, \frac{\delta_{ij} k^2 - k_ik_j}{|k|}  \, a\dgg_j(k)
   \weak \hbar \, |k|  \, a\dgg_i(k),
\eeq
because $k_j\op a\dgg_j(k)=k_j \op E_j(k)\weak0$. The correct commutator
relation holds only weakly, but this is enough to get the right eigenvalue
equations for the physical states. Hence, we finally arrived at the correct
quantization of the free electro-magnetic field. Everything has been more or
less straightforward, we just had to apply the rules of the Dirac programme. I
should emphasize that, in contrast to other quantization schemes, our final
result is completely {\em and manifestly\/} gauge covariant. At no point we had
to impose a gauge fixing. All our stated are represented by wave functionals,
which can be obtained by acting repeatedly with the differential operators
corresponding to $\op a\dgg_i$ on the vacuum functional \eref{qed-vac}, and all
these functionals are gauge invariant.

The problem with the wave functionals is, beside the fact that they will become
rather awkward objects after after a while, that we do not have any idea of how
to represent the scalar product. It should be some functional integral which
becomes finite exactly for the states in our Fock space. It is quite obvious
that then we are confronted, as already mentioned, with all the problems of the
path integral approach. We had to integrate over gauge degrees of freedom,
which requires some gauge fixing, ghosts, etc. So, it is much more suitable to
take the Fock basis and forget the wave functionals. All we need is the
representation of the Hamiltonian in terms of the $\op a_i$ and $\op a\dgg_i$
operators, and their commutators. This already defines them uniquely as
\eref{qed-a-a}. Of course, this was the way the special wave functional
\eref{qed-vac} was found: by solving the differential equation
\eref{qed-vac-eq}, just like one does it to find the ground state of the
harmonic oscillator in the ``$x$-representation''.

\subsection*{The parameter time}
We shall now consider another class of systems with gauge degrees of freedom
which are quite different from the standard ones. Their common feature is that
they possess an invariance under the diffeomorphism group of the underlying
spacetime, which physicist often express as {\em general covariance}. Typical
examples are the relativistic point particle, string theory, and general
relativity. For the first, the gauge group consists of the diffeomorphisms of
the real line, represented by ``reparametrization'' of the world line of the
particle. The same in one dimension more holds for the string, and in gravity
the gauge group can be considered as the diffeomorphism group of spacetime.

As the first step in the Dirac programme is to write the action as an integral
of a Lagrangian over time, we are confronted with the problem that the natural
action for these systems is not given in that form. For the relativistic
particle or string it is given as an integral over the worldline or world
surface instead, and in gravity time itself becomes a dynamical object. The way
out in these cases is to consider some ``time-like'' parameter, which behaves
{\em almost\/} like a time coordinate, and perform the Dirac programme using
this {\em parameter time}. Hence, all we need is that the action is given in
the form of an integral of some Lagrangian over a real parameter. For the
relativistic particle this can be chosen to be any parameter on the world line,
for the string the non-compact coordinate on the world surface, and for gravity
we need a {\em foliation\/} of spacetime, or equivalently there must be a
special global coordinate $\tau\in\RR$ such that spacetime becomes a direct
product of that $\RR$ with some space manifold. In all these cases the action
integral can be split into a ``spatial'' integral and a ``time'' integral,
which is similar to the split we had to make in field theory, except that out
time is not the physical time. The Lagrangian then takes the usual form, it
becomes a function of the fields and there velocities, which are now defined
with respect to the parameter time.

After deriving the Lagrangian, we can simply forget that the parameter time is
not the physical time, and perform the Dirac programme straightforwardly. The
reparametrization invariance will show up as a gauge degree of freedom, i.e.\
as a set of first class constraints. These will take care that the state as
well as the observables become gauge-invariant objects, which are independent
of the chosen time coordinate. Moreover, all these theories have another common
feature which reflects the fact that the ``canonical time'' is not the physical
time: the Hamiltonian will always vanish on the constraint surface. This has
two consequences. It means that there is no energy ``conjugate'' to the
parameter time, which would be unphysical too. It also means that the total
Hamiltonian can be written as a linear combination of constraints, hence as a
generator of a gauge transformation. This again means that time evolution
becomes a pure gauge transformation, which is what we have to expect because
any type of time reparametrizations belong to the gauge group. From the
physical point of view, there will be no time evolution at all. How can this
be? We know that the relativistic particle or string does evolve in time, and
in gravity spacetime is also dynamic. But we have to keep apart the two things
called time, the physical time and the parameter time, and if we do it
carefully, we will see that there is no contradiction.

\subsection*{The relativistic point particle}
All this is most simply explained if we consider the relativistic point
particle as an example. The action of a particle of mass $m$ moving in flat
Minkowski space is given by the length of the path. To make it not too simple,
let us couple the particle to an external electro-magnetic field, then the
action becomes
\beq[r-act]
   \act[q(t)] = \intdtau   m \sqrt{-\dt q^\mu(\tau) \dt q_\mu(\tau)}
                                - e \, \dt q^\mu(\tau)\,A_\mu(\q(\tau)),
\eeq
where $\mu,\nu=0,1,2,3$, and the metric again $g_{\mu\nu}=\diag(-1,1,1,1)$. The
extrema of this action are of course the timelike straight lines, but on these
lines the parameter $\tau$ can be chosen quite arbitrary. We can replace it by
any function $\tau\mapsto f(\tau)$ and we still get the same physical solution.
The gauge group is therefore expected to be the diffeomorphism group of $\RR$.
Let us apply the method we have developed. The Lagrangian is
\beq[r-lag]
  \lag[\q,\v] = m \sqrt{-v^\mu v_\mu} - e \, v^\mu\,A_\mu(\q) .
\eeq
For the momenta, we find
\beq[r-mom]
  p_\mu = \deldel \lag / v^\mu / = -\frac{m v_\mu}{\sqrt{-v^\nu v_\nu}}
               - e\, A_\mu,
\eeq
and this gives a primary constraint
\beq
  \con = \ft12\big((p_\mu + e \, A_\mu)(p^\mu + e \, A^\mu) + m^2 ).
\eeq
Now let us consider the Hamiltonian. If our system is gauge-invariant under
arbitrary rescaling of the time coordinate, then the Lagrangian must be a
homogeneous function of the velocities. This can be see as follows. If the
action is invariant under rescaling $\tau \mapsto \tau'$, then we must have
\beq
  \d\tau \, \lag(\q,\v) = \d\tau' \, \lag(\q,\v')  =
  \d\tau' \, \lag(\q,\dd \tau / \tau' / \v ) ,
\eeq
because the velocities scale as $\v\,\d\tau=\v'\,\d\tau'$. So $\lag$ must be
homogeneous in $\v$, which is obviously true for our example. The function
$f=\p\cdot\v-\lag$, which has to be extremized to give the Hamiltonian, is
homogeneous as well. For such a function we have $\v\cdot\del f/\del\v=f$. This
tells us that whenever it has an extremum, the value of that extremum is zero.
So the Hamiltonian will always vanish on the constraint surface. As already
mentioned, this is in agreement with the physical statement that there can be
no conserved charge, i.e.\ no ``energy'', conjugate to an unphysical time. The
total Hamiltonian is therefore just a ``linear combination'' of the constraint
\beq[r-ham]
  \ham = u \, \con=
       \ft12 u \, \big( (p_\mu + e \, A_\mu)(p^\mu + e \, A^\mu) + m^2 \big),
\eeq
with a free parameter $u$. There are no more constraints as $\con$ clearly
commutes with $\ham$, and for the same reason it is first class. To derive the
evolution equations, we have to take into account that $A_\mu$ depends on $\q$:
\beq[r-evol]
  \dt q^\mu = \pois{q^\mu}{\ham} \weak u \,(p^\mu + e\, A^\mu) , \quad
  \dt p_\mu = \pois{p_\mu}{\ham} \weak
                     - u \,(p^\nu + e\, A^\nu) \, e \,\del_\mu A_\nu.
\eeq
You can check that, if $\p$ is eliminated from these equations, the potential
enters the resulting second order equation for $\q$ only via the field
strength.

Let us take $A_\mu=0$ for the moment. What are the gauge transformations
generated by $\con$? The brackets are
\beq[r-gauge]
 \pois{q^\mu}\con = p^\mu , \qquad
 \pois{p^\mu}\con = 0.
\eeq
This is a displacement in the configuration variable $\q$ along the direction
of $\p$. Note that on the constraint surface (which is the direct product of
Minkowski space with the two parts of the mass shell) $\p$ is always a nonzero
timelike vector. We find that two states $(\q,\p)$ and $(\q',\p')$ are
physically equivalent, if $\p=\p'$ and $\q-\q'\propto\p$. Hence, if we make a
gauge transformation, the particle will appear somewhere else in spacetime, at
a place where is ``was'' or ``will be'' at an earlier or later ``time'', but it
will have the same momentum. If we look at the class of all states that are
equivalent to some given state, we find that it is a straight line in phase
space: the value of $\p$ lies on the mass shell, is constant, and gives the
direction of the line in the configuration space, which is Minkowski space. So,
the {\em world line\/} in phase space that passes through a given state
$(\q,\p)$ is the {\em equivalence class} of that state. The same holds for the
particle in a background field. The states that are equivalent to some initial
state $(\q,\p)$ are exactly the points on the world line of a particle moving
in the background field and passing though the event $\q$ with momentum $\p$.
Remember that we defined physical states to be these gauge-equivalence classes,
so the physical states in our case are the world lines in phase space.

Let us consider this from a different point of view and look at the {\em time
evolution\/} of a state on the constraint surface. We said that we are free to
choose the parameter $u$ in the Hamiltonian as we like. So let us take $u=0$.
Then there is no time evolution at all. Given an initial state $(\q_0,\p_0)$,
at $\tau=0$, integrating the Hamilton equations gives $\q(\tau)=\q_0$ and
$\p(\tau)=\p_0$. This is a quite funny solution, the particle doesn't move
through spacetime, it just sits at one event. Seems that it doesn't make any
sense to consider such solutions. But you can convince yourself that this is a
solution that extremizes the action. We could avoid such silly things by
imposing boundary conditions on the paths like $q^0\to\pm\infty$ for
$\tau\to\pm\infty$. But that doesn't make sense in the Hamiltonian picture.
Here, were we are only dealing with states that evolve ``locally'' in time and
not with the whole time evolution. We can generate even more silly solutions:
by taking $u$ to be an oscillating function, the particle will move up and down
along the world line, or we can make it running backwards in time (don't mix
this up with replacing $\p\mapsto -\p$ which is not a gauge transformations and
not even a symmetry if $A_\mu\neq0$; we'll come to this in a moment).

But we have to remember that all these ``evolutions'' are just gauge
transformations. There will never be any {\em physical} time evolution, as all
the states we pass through will always correspond to the same physical state,
the same equivalence class of states. Nevertheless, somehow the particle {\em
should\/} move through spacetime and we should be able {\em see\/} it moving.
In other words, we should be able to {\em observe\/} the particle moving
through spacetime. We have to remember what we said about {\em observables}. An
observable is a phase space function that is gauge-invariant, or has weakly
vanishing brackets with the constraints. For our particle, we must have
$\{\obs,\con\}\weak0$. Without the exterior field, the $p_\mu$ are observables,
but they are not all independent, because we have $\p^2\weak m^2$. So they make
up three independent real observables $P_i=p_i$, $i=1,2,3$, and one sign $C$
which gives $p^0 = C \sqrt{p_ip_i+m^2}$ and tells us on which part of the mass
shell we are. Of course, $C$ is an observable even if the field is switched on,
and by inspecting the equations of motion, we find that it is the sign of the
charge of the particle. At this point it is really important to note that the
transformation $C\mapsto -C$ has {\em nothing\/} to do with and cannot be
compensated by choosing another gauge, i.e.\ changing the value of $u$. It
really changes the physical state, as the particle is now oppositely charged
and moves on a totally different line, even if it is given the same 3-momentum
at the same spacetime point initially. And note also that this ``double'' phase
space occurs already at the classical level. The action \eref{r-act} has two
types of solutions: electrons and positrons, which behave different if there is
an external field. This is not a quantum-effect, as is sometimes stated. The
discrete observable $C$ is a result of the classical Hamiltonian formalism.

Again without the field, the observables $P_i$ and $C$ are also conserved
charges as we should expect. But what was the difference between observables
and conserved charges? In addition to the first class constraints, conserved
charges must have vanishing brackets with the Hamiltonian. But this doesn't
give anything new here: the Hamiltonian consists of nothing but the constraint.
Here we found another general property of systems with parameter time. Every
observable automatically becomes a conserved charge. This is not really
surprising, because if the time parameter $\tau$ is not the physical time and
can be reparametrized arbitrarily, we should expect that a quantity that has a
physical meaning must not depend on $\tau$. And ``conserved'' in this context
means with respect to $\tau$ and not with respect to the {\em physical\/} time,
which in our case may be taken to be the phase space function $q^0$ (which is
not an observable!).

Beside the momenta, which are also ``conserved'' with respect to the physical
time (whatever that in this context might mean), are there other observables?
Let us think in a more physical way: what else can we measure? For example,
where {\em in space\/} is the particle at the {\em physical\/} time $q^0=t$?
Given a state $(\q,\p)$, how can we find the answer, if not accidentally
$q^0=t$. In that case, the answer is of course that the particle is at the
space point with coordinates $q_i$. Here is the general answer: make a gauge
transformation from $(\q,\p)$ to $(\q',\p')$ such that $q^{\prime 0}=t$. By
definition, this does not change any physical quantity. Then $q'_i$ gives the
space point we are looking for. For $A_\mu=0$, it is not too difficult to write
down an observable for this measurement:
\beq[r-q-t]
  Q_i(t) \weak q_i + \frac{(t - q^0)}{p^0} p_i .
\eeq
This gives a different {\em phase space function\/} for each value of $t$,
which for any state on the constraint surface gives the space point where the
particle is, was, or will be when $q^0=t$. It is well defined because
$p^0\neq0$ on the surface. You can check that its bracket with the constraint
vanishes, so it is a conserved charge. We can also switch to another observer
and ask for the position he will find at some time $t'$ in his reference frame.
Clearly, an explicit expression for this can be given by Lorentz-transforming
\eref{r-q-t} suitably, or more geometrically by finding the intersection point
of the world line with some spacelike hypersurface. In principle, we can
express everything which is a property of the world line itself and not of its
parametrization in form of an observable or conserved charge. Note again that
$t$ has {\em nothing\/} to do with the parameter $\tau$ and ``conserved'' does
not mean that $Q_i(t)$ does not depend on $t$. If we set $Q_i=Q_i(0)$, we get
the following complete set of independent observables:
\beq[r-obs-0]
  C = \sign (p^0) , \qquad
  P_i = p_i , \qquad
  Q_i = q_i - \frac{q_0}{p_0} p_i .
\eeq
The more familiar ones are functions thereof, e.g.\ the angular momenta
\beq[r-j]
  J_{\mu\nu} = q_\mu p_\nu - q_\nu p_\mu \follows
  J_{0 i} = C Q_i \sqrt{P_iP_i+m^2} , \quad
  J_{ij} = Q_i P_j - Q_j P_i,
\eeq
which are not independent. We see that to extract physical properties of a
state, we have to refer to the physical states, i.e.\ the equivalence classes
under gauge transformations, which are the world lines. If we do this, we can
totally ignore any evolution generated by the Hamiltonian, as there is none.
Instead, to see a physical evolution, we must refer to the physical time, which
in this case can easily be identified with a special phase space function
$q^0$. We can then ask for {\em correlations\/} between this physical time and
other phase space functions, and it is this what gives us observables.

In principle, we can do the same if there is a background field, things just
get technically more involved. What we still have is the following. For every
equivalence class and every physical time $t$ there is exactly one
representative for which $q^0=t$. This is because the world line is always
timelike. Now consider an arbitrary phase space function $F$, not necessarily
an observable. For a given physical state, the value of $F$ will depend on the
representative. But there is an {\em observable\/} $F_t$ which is defined to be
the value of $F$ at that representative where $q^0=t$. Clearly, the physical
interpretation of that observable is the value of $F$ {\em at the time\/} $t$.
By definition $F_t$ is constant along the world line, so it will have a
vanishing bracket with the constraint. If we take $F=q_i$, which is not an
observable, then $F_t$ will be the space point where the particle is at
$q^0=t$, and this is the observable $Q_i(t)$ as defined above. Of course, to
give an explicit phase space functions in general one needs to solve the
equations of motion. This is what we could do for the field-free case above,
which gave us the rather simple expression \eref{r-q-t} for $Q_i(t)$.

Hence, we found that the main problem for systems with parameter time is the
construction of observables. It is somehow clear from the physical point of
view, as all we have to do is to identify the physical parameters describing
the state, but it is rather complicated to give explicit representations of
these functions in terms of phase space function. In more complicated theories
like gravity this is indeed one of the major problems. But the ansatz is very
similar: one has to identify some phase space function which describes the
physical time, and then consider correlations with other phase space functions.
What makes this a bit harder than for our simple system is that in gravity
there is no ``global'' physical time. So one either has to deal with a local
time, which can be described by some kind of ``clock field'' and refer to this,
or one can use ``cosmological clocks'', which in principle is the translation
of ``boundary conditions'' from mathematical into physical language.

Let us now quantize the relativistic particle. We take wave functions
$\Psi(\q)$ to represent states $\kt\Psi$, and the operators are
\beq
  \op q_\mu \, \Psi(\q) =  q_\mu \Psi(\q), \qquad
  \op p_\mu \, \Psi(\q) = - \i \hbar \deldel \Psi / q^\mu / (\q).
\eeq
The physical states are the solutions of the Klein Gordon equation (with
$A_\mu=0$ from now on)
\beq
   \op \con \ket \Psi = 0 \equiv
  \Big( \deldel / q^\mu / \deldel / q_\mu / -  \frac{m^2}{\hbar^2} \Big)
                 \Psi(\q)=0.
\eeq
We know the complete set of solutions, and we can most conveniently take the
$\op \p$ eigenstates as a basis for the physical state space. For each
three-vector $ k$ and an additional sign $c=\pm1$ we have a state $\kt{ k,c}$
which is represented by the wave functional
\beq
   \Psi_{ k,c}(\q) =
       \exp\big(\i k_i q_i - \i c \,\omega q^0  ),
      \txt{with} \omega = \sqrt{k_ik_i+ (m/\hbar)^2}.
\eeq
They provide the eigensystem of the observables $P_i$ and $C$, which form a
maximally commuting subset of \eref{r-obs-0}:
\beq
  \op P_i \ket{ k,c} = \hbar \, k_i \ket{ k,c}, \qquad
  \op C \ket{ k,c} = c \ket{ k,c}.
\eeq
There is no time evolution at the quantum level, for the same reason as above:
a physical state is a gauge invariant object, and time evolution is nothing but
a gauge transformation. Here it is even more apparent that we really need the
observables to extract physical information. At the classical level we could
answer the question ``where is the particle at physical time $q^0=t$?'' by
looking at the equivalence classes, the world lines of the particle. We gave an
answer by choosing a special representative $(\q,\p)$ with $q^0=t$. We cannot
do this here any more. Quantum physical states are gauge invariant objects by
themselves, we cannot choose a particular representation. Moreover, we cannot
reinterpret the wave function on spacetime as a {\em spatial\/} wave function
$\Psi(q_i)$ which evolves in time $q^0$, as is sometimes suggested. This causes
lots of problems, as for example there is no ``conservation of probability'',
so one cannot define suitable expectations values etc. The only way out usually
suggested is that one has to consider pair creation an annihilation, and go
over to multi-particle physics.

But this is not true. It is sufficient to note that the Klein Gordon equation
is {\em not\/} a ``relativistic Schr\"odinger equation''. It is {\em not\/} a
time evolution equation for the state but a {\em constraint}. We shall see that
if we keep this in mind and find a suitable scalar product, we will not have
any problems. We can quantize one single relativistic point particle. The only
way to extract physical information is via the expectation values of
observables..
\beq
  \expect{Q_i(t)} = \bra \Psi \op Q_i(t) \ket \Psi
\eeq
will gives us the expected position of the particle at physical time $t$, as
seen by the observer whose restframe is the one we are working in. A similar
expression can of course be given for any observer by replacing the operator
with a Lorentz transformed one, just like in the classical case.\footnote{Here
we are ignoring the problem that in relativity we cannot really make such kind
of measurements, as they would involve a whole spacelike hypersurface as
``measuring device''. But you can think of equivalent ``local'' measurements
where the observables are given as projectors onto some eigenstates of
$Q_i(t)$.}

But what is the scalar product? Clearly, we want that the observables
\eref{r-obs-0} become Hermitian, and this will fix the product up to
normalization. The momenta $P_i$ and the charge $C$ become Hermitian if their
eigenstates are orthogonal. Hence, we must have
\beq
  \braket { k,c}{ l,c'}
    = (2\pi)^3 \, \delta_{c,c'} \, \delta^3( k -  l) \, f( k),
\eeq
for some positive function $f( k)$. To fix this function, we have to require
$Q_i$ to be Hermitian. It was given by $Q_i=q_i-(p_0)^{-1}q_0 p_i$. This causes
a problem, as it is not clear in which order the non-commuting
operators\footnote{Note that $(\op p_0){}^{-1}$ is a well defined operator on
the physical state space} $(\op p_0){}^{-1}$ and $\op q_0$ are to be put. Let
us choose the most general possibility, which is
\beq
  \op Q_i&=&\op q_i - (1-\alpha) \, \op q_0 \, (\op p_0)^{-1} \op p_i
                    - \alpha     \, (\op p_0)^{-1} \, \op q_0 \,\op p_i \zl
         &=&\op q_i - \, \op q_0 \, (\op p_0)^{-1} \op p_i
                    + \i \alpha \hbar \, (\op p_0)^{-2}\, \op p_i.
\eeq
Here we used the commutator $[(\op p_0)^{-1},\op q_0]=-\i\hbar (\op p_0)^{-2}$,
showing explicitly that the ordering ambiguity corresponds to an $O(\hbar)$
term. As this is an observable (for any value of $\alpha$), its action on a
physical state should be a physical state again. In fact, one finds that
\beq
  \op Q_i \ket{ k,c} =
   -\i\deldel / k_i / \ket{ k,c}
     + \i \alpha\, k_i \omega^{-2} \ket { k,c}.
\eeq
This gives the following matrix element for $\op Q_i$:
\beq
  &&\bra { l,c} \op Q_i \ket{ k,c} =
   - \i (2\pi)^3 \Big( \, \del_i
             \big( \delta( k- l) \, f( k) \big ) +
       \delta( k- l) \,
        \alpha \, k_i \omega^{-2}  \, f( k) \Big).
\eeq
It becomes Hermitian if
\beq
    \del_i f( k) = 2 \alpha \,  k_i\omega^{-2} \, f( k)
   \follows
     f( k) = \omega^{2\alpha}.
\eeq
This shows how requiring the real observables to be Hermitian gives a {\em
relations\/} between the representation map $\op{{\ }}$ and the scalar product,
as the parameter $\alpha$ is still free, and for any value we choose we will
find $\op Q_i$ to be Hermitian. The same, of course, also holds for the
operators $\op Q_i(t)$ for the place of the particle at $q^0=t$, as they are
given as a sum of $\op Q_i$ and another Hermitian operator. But nevertheless
there is a preferred choice for the product that fixes $\alpha$. Consider the
angular momentum operators introduced in \eref{r-j}. As $\op Q_i$ does not
commute with $\op P_i$, there is an ordering ambiguity in the definition of
$\op J_{0i}$. If we want it to be Hermitian, there is only one possible
ordering, which is the symmetrized product. We can also turn the argument
around and consider the classical equation $Q_i=(p_0)^{-1} J_{i0}$. This can
only become a relation between Hermitian operators if we take the symmetrized
product in the corresponding quantum relation $2 \, \op Q_i=(\op p_0)^{-1} \op
J_{i0} + \op J_{i0} (\op p_0)^{-1}$. Now we impose an {\em additional\/}
requirement. We want that the angular momentum is represented in the standard
form $\op J_{i0}=\op q_i\op p_0-\op q_0 \op p_i$, where no ordering ambiguities
occur. You can easily see that this fixes $\alpha=1/2$.

Note, however, that this is just a convenient choice, it does not have any
impact on the physical properties of our system, it just provides the nicest
possible representation. Otherwise we would have to work with slightly
unconventional angular momentum operators. We can also see that $\alpha=1/2$ is
a ``canonical'' choice, because then the product becomes
\beq
  \braket { k,c}{ l,c'}
    = (2\pi)^3 \, \delta_{c,c'} \, \delta^3( k -  l)
     \, \omega( k),
\eeq
and this is the unique Lorentz invariant delta function on the mass shell.
Clearly, it would have been much easier to require the angular momentum
operators to be Hermitian straight away, which more or less immediately tells
us that the delta function in the product must be Lorentz invariant. But for
more complicated theories it might not be so obvious that there is such a
distinguished representation. And it is important to note that there are some
ambiguities in the scalar product and the representation which cannot be fixed
by physical arguments.

\subsection*{String theory}
The same technique can be applied to string theory. It is very much like the
point particle, except that there is one dimension more. The action of a string
moving in a flat Minkowski space is given by the area of its world surface. The
latter is given as a function $q^\mu(\sigma,\tau)$ of two parameters. We
consider it as a two dimensional field theory. The ``time coordinate'' will be
$\tau$, and $\sigma$ is considered as a spatial coordinate, which runs from $0$
to $2\pi$. We choose the closed string here, i.e.\
$q^\mu(0,\tau)=q^\mu(2\pi,\tau)$. The action can then be written as
\beq[s-act]
  \act[\q] = \int\d\tau\,\d\sigma \,\,
             \sqrt{ \del_\sigma  q_\mu  \del_\tau   q^\mu \,
                    \del_\sigma  q_\nu  \del_\tau   q^\nu  -
                    \del_\sigma  q_\mu  \del_\sigma q^\mu \,
                    \del_\tau    q_\nu  \del_\tau   q^\nu }.
\eeq
The expression under the square root is the negative determinant of the metric
on the world surface that is induced by the embedding of the world surface into
the background space, so this will give the total area covered by the surface.
By splitting off the ``time'' integral, we get the Lagrangian
\beq[s-lag]
  \lag[\q,\v] = \intdsig
             \sqrt{ \qq_\mu  v^\mu \, \qq_\nu  v^\nu  -
                     v_\mu  v^\mu \, \qq_\nu  \qq^\nu  },
\eeq
where the prime denotes the derivative with respect to $\sigma$.
Similar to the relativistic particle, this is homogeneous of degree one in the
velocities, showing that the time $\tau$ is an unphysical parameter only. For
the momenta we find
\beq[s-mom]
  p_\mu = \deltadelta \lag /  v^\mu / =
             \frac{ \qq_\mu \, \qq_\nu  v^\nu  -
                    v_\mu \, \qq_\nu  \qq^\nu }{
             \sqrt{ \qq_\mu  v^\mu \, \qq_\nu  v^\nu  -
                     v_\mu  v^\mu \, \qq_\nu  \qq^\nu  }},
\eeq
and the Poisson brackets are
\beq[s-pois]
  \pois{q_\mu(\sigma)}{p_\nu(\rho)}=g_{\mu\nu}\delta(\sigma-\rho).
\eeq
And we get the following set of primary constraints:
\beq[s-c12]
  \fcl_1(\sigma) =  \ft12 p_\mu(\sigma) \, p^\mu(\sigma) +
                    \ft12 \qq_\mu(\sigma) \, \qq^\mu(\sigma), \qquad
  \fcl_2(\sigma) =  p_\mu(\sigma) \, \qq^\mu(\sigma) .
\eeq
We will shortly see that these are first class and that there are no secondary
constraints, but before showing this let us briefly look at the gauge
transformations they generate, as this is typical for every general covariant
theory. The same type of constraints are found in gravity as well.
We can define the ``linear combinations''
\beq
  \fcl_1[u] = \intdsig u(\sigma)\, \fcl_1(\sigma) \qquad
  \fcl_2[v] = \intdsig v(\sigma)\, \fcl_2(\sigma).
\eeq
The gauge transformations generated by them are
\beq
   &&\pois{q_\mu}{\fcl_1[u]} =  u \, p_\mu, \qquad
     \pois{p_\mu}{\fcl_1[u]} =  \del_\sigma ( u \, \del_\sigma q_\mu ), \zl
   &&\pois{q_\mu}{\fcl_2[v]} =  v \, \del_\sigma q_\mu, \qquad
     \pois{p_\mu}{\fcl_2[v]} =  \del_\sigma (v \, p_\mu).
\eeq
Let us look at the second constraint first. The transformations on the right
hand side are exactly those corresponding to an infinitesimal shift $\sigma
\mapsto \sigma + v$, where $v$ is some ``vector field'' on the string. Thereby,
the coordinate behaves like a scalar, and the momentum like a density of weight
one. The gauge symmetries generated by the second constraint are the
diffeomorphisms of the spatial coordinate of the string. A similar {\em
diffeomorphism constraint\/} is found in all general covariant theories. In
gravity we have three of them at each space point, generating the
diffeomorphism group of the spatial manifold. We didn't have this for the point
particle because there the ``spatial manifold'' was just a point. It is quite
easy to give the full set of phase space functions that weakly commute with
that constraint. They are exactly those which are independent of the special
parametrization $\sigma$. This is very similar to the electro-magnetic field,
and these {\em kinematical constraints\/} are in general easy to solve. We just
have to identify the invariants of a rather simple gauge group. Using the
standard representation, this also holds for the quantum theory. The wave
functional can be any invariant that does not depend on the momenta.

The other constraint, however, behaves quite differently. It looks similar and
the transformations generated are also similar to those of the point particle.
It generates the diffeomorphisms of the $\tau$ coordinate, but because this is
our canonical time coordinate, these diffeomorphisms are {\em dynamical}. They
are not realized by simply replacing $\tau\mapsto\tau+u$, which doesn't make
sense in the Hamiltonian picture. Instead, the contraint generates those
transformation on the phase space that formally corresponds to the time
evolution. It replaces the (non-existing) time evolution generated by the
Hamiltonian, and is therefore also called {\em Hamiltonian constraint} or {\em
dynamical constraint}. We had the same for the point particle: the constraint
generated a gauge transformation which effectively looks like a time evolution.
Here the situation is just a bit more complicated as we do not only have one
dynamical constraint, but one for each $\sigma$. The typical structure of the
dynamical constraint is the quadratic term in the momenta, which we also had
for the point particle and which also appears in gravity, where the (quantized)
dynamical constraint is the famous {\em Wheeler De~Witt equation\/}.
Because of this quadratic term, it no longer generates gauge transformation on
the wave functional, like the kinematical constraints did. So it is much harder
to solve. It contains the whole infomation about the dynamics of the system, as
there is no Hamiltonian.

Hence, we find that generally in systems with parameter time the dynamics is
encoded in the constraints and not in the energy function, which always
vanishes. These dynamical constraints are typically quadratic in the momenta,
like the energy for unconstrained systems. For field theories like strings or
gravity this causes another problem. In the standard representation the square
of the momentum operator will not be well defined, as it becomes a functional
differential operator. This problem was present in electro-dynamics as well but
there it shows up only in the Hamiltonian and not in the constraints, so we
could simply ignore it until we finally ``automatically'' found an expression
for the Hamiltonian that was well defined on the physical state
space.\footnote{The functional derivative operator inside the integral
\eref{qed-ham} is not well defined when acting on a general non-physical
functional of the gauge potential, try it with the Chern Simons form! Note that
the problem is not that the integral doesn't converge, the operator itself is
ill-defined.} This of course leads to the well known problems of quantum field
theory, we need a regularization etc. All these typical problems will now
already show up in the constraint algebra.

So before quantizing the string we should think about a regularization, as it
is quite obvious that there is no well-defined quantum operator for the
constraint $\fcl_1$, which contains the square of the momentum. The most
suitable regularization is to Fourier transform the $\sigma$-coordinate, which
is compact and therefore we can replace integrals by sums. Then we can take the
inifinit sums quite naturally as the ``regulator~$\to0$'' limits. But before we
do this, let us first write the constraints in a more convenient way. And, of
course, we  still have to prove that there are no secondary constraints. As the
Hamiltonian vanishes weakly, all we have to do is to check whether the brackets
of the primary constraints with themselves already vanish weakly.
We can write the constraints more symmetrically by introducing
\beq[s-ab]
  a_\mu = \ft12 ( p_\mu + \qq_\mu ) , \qquad
  b_\mu = \ft12 ( p_\mu - \qq_\mu )
\eeq
as auxilliary variables. Their brackets are
\beq
 && \pois {a_\mu(\sigma)}{a_\nu(\rho)} =
   \ft12 g_{\mu\nu} \, \delta'(\sigma-\rho),   \zl
 && \pois {b_\mu(\sigma)}{b_\nu(\rho)} = -
   \ft12 g_{\mu\nu} \, \delta'(\sigma-\rho),   \qquad
  \pois {a_\mu(\sigma)}{b_\nu(\rho)} = 0.
\eeq
You can think of these two sets of variables as describing ``right'' and
``left'' moving modes on the string world surface. They form an {\em almost\/}
complete set of variables. They are not completely independent, as
\beq[s-ab-dep]
   P_\mu =  \intdsig p_\mu(\sigma)
        = 2 \intdsig a_\mu(\sigma) = 2 \intdsig b_\mu(\sigma)
\eeq
is the {\em total momentum}, and the only quantity that is missing in these
variables is the {\em center of mass\/} of the string
\beq
   Q_\mu = \frac 1{2\pi} \intdsig q_\mu(\sigma).
\eeq
But this does not show up in the constraints, so that they can be combined into
two simple functions of the $a$'s and $b$'s:
\beq[s-cpm]
  \fcl_+ = \ft12 (\fcl_1 + \fcl_2) = a_\mu \, a^\mu ,\qquad
  \fcl_- = \ft12 (\fcl_1 - \fcl_2) = b_\mu \, b^\mu .
\eeq
For them, we find
\beq[s-c-pois]
  \pois{ \fcl_+(\sigma) }{ \fcl_+(\rho) } &=&
   \delta'(\sigma-\rho) \, \big( \fcl_+(\sigma) + \fcl_+(\rho) \big), \zl
  \pois{ \fcl_-(\sigma) }{ \fcl_-(\rho) } &=& -
   \delta'(\sigma-\rho) \, \big( \fcl_+(\sigma) + \fcl_-(\rho) \big),
\eeq
and the mixed bracket vanishes. So, we don't get anything new. We have two
separate algebras of first class constraints. A more familiar form of the
algebra is obtained by Fourier transforming everything which is a function of
$\sigma$. For all such functions we define
\beq[s-trafo]
  f(k) = \intdsig e^{-\i k\sigma} \, f(\sigma) \equiv
  f(\sigma) = \sumk e^{\i k \sigma} \, f(k),
\eeq
where $k$ is an integer. The $k=0$ modes of the canonical variables then become
the center of mass and the total momentum
\beq
 Q_\mu=\frac {q_\mu(0)}{2\pi}  , \qquad
 P_\mu=p_\mu(0) = 2 \, a_\mu(0) = 2 \, b_\mu(0)
\eeq
The basic Poisson brackets are
\beq[s-p-q-com]
   \pois{ q_\mu(k)}{ p_\nu(l) } = 2 \pi \, g_{\mu\nu} \, \delta_{k+l}
    \follows \pois{Q_\mu}{P_\nu} = g_{\mu\nu}.
\eeq
for the canonical variables, and
\beq[s-a-b-com]
  \pois{ a_\mu(k) }{ a_\nu(l) } =
      \i\pi \, k \, g_{\mu\nu} \, \delta_{k+l}, \qquad
  \pois{ b_\mu(k) }{ b_\nu(l) } =
    - \i\pi \, k \, g_{\mu\nu} \, \delta_{k+l},
\eeq
where $\delta_k=1$ for $k=0$ and $0$ otherwise. For the transformed constraints
we now get the well known Virasoro algebra
\beq[s-vira]
  \pois{ \fcl_+(k) }{ \fcl_+(l) } &=& \i \, (k-l) \, \fcl_+(k+l), \zl
  \pois{ \fcl_-(k) }{ \fcl_-(l) } &=& -\i \, (k-l) \, \fcl_-(k+l).
\eeq
To quantize the string, we choose the standard representation again. We
represent a quantum state by a wave functional $\Psi[\q]$, and the basic
operators are
\beq
   \op q_\mu(k) \, \Psi[\q] = q_\mu(k) \Psi[\q] , \qquad
   \op p_\mu(k) \, \Psi[\q] = -2\pi\i\hbar \, \deldel \Psi / q^\mu(-k) / [\q].
\eeq
The operators for the $a$'s and $b$'s are straightforwardly obtained by using
the transformed of \eref{s-ab}:
\beq[s-ab-four]
   \op a_\mu(k) = \ft12\big(\op p_\mu(k) + \i k \, \op q_\mu(k) \big), \qquad
   \op b_\mu(k) = \ft12\big(\op p_\mu(k) - \i k \, \op q_\mu(k) \big).
\eeq
Finally, the Fourier transform of the constraints gives the following quantum
representation:
\beq[s-const]
  \op\fcl_+(k) = \suml \op a_\mu(l) \, \op a^\mu(k-l) , \qquad
  \op\fcl_-(k) = \suml \op b_\mu(l) \, \op b^\mu(k-l).
\eeq
If we look at the products of operators appearing therein, we find that for
$k\neq0$ we always multiply commuting operators, so there is no ambiguity when
quantizing those. However, if $k=0$, each term in the sum consists of
non-commuting operators, and it is not clear in which order they have to
appear. Let us rewrite the relevant constraint as
\beq
  \op\fcl_+(0) = \frac 1{8\pi} \, \op P_\mu \, \op P^\mu +
                 \sumll \op a_\mu(l) \, \op a^\mu(-l) +
                 \sumll \op a_\mu(-l) \, \op a^\mu(l)  ,
\eeq
and similarly for $\fcl_-(0)$. I just wrote the sum over the positive and
negative $l$'s separately, and replaced $a_\mu(0)$ by the total momentum. In
this form, there will never be a solution to that constraint, simply because
the sums will never converge both when acting on some state. If the first
converges when applied to a state $\kt\Psi$, we must have
\beq
  \op a_\mu(l) \, \op a^\mu(-l) \ket\Psi \to 0 \txt{for} l \to \infty.
\eeq
Using the commutator of the $a$'s this implies
\beq
  \op a_\mu(-l) \, \op a_\mu(l) \ket\Psi - \pi \hbar \, D\, l \ket \Psi\to 0
  \txt{for} l \to \infty,
\eeq
where $D$ is the dimension of the target spacetime. The second term will never
converge, and so the second sum above cannot converge either. We have to
reorder the operators such that both sums can converge. The condition for this
is that {\em almost all\/} terms appear in the same order in both sums. But
which order shall we choose? To see that there is a canonical choice, consider
the following commutators of complex conjugate operators
\beq[s-k-k-com]
  \comm {\op a_\mu(-k) }{ \op a_\nu(k) } = \pi\hbar \, k \, g_{\mu\nu} , \qquad
  \comm {\op b_\mu(k) }{ \op b_\nu(-k) } = \pi\hbar \, k \, g_{\mu\nu} .
\eeq
They behave like annihilation and creation operators. For them, this was
something we also found for the electro-magnetic field, it is important that
the commutator of an annihilation operator with a creation operator is
positive. Otherwise we cannot define a proper scalar product. The Lorentzian
signature of the metric is a bit problematic here, but let us just ignore this
for a moment and assume we are in a Euclidian space. Then we must conclude
that, if we want to build up a Fock space using these operators, we must take
$\op a_\mu(-k)$ and $\op b_\mu(k)$, for $k>0$, to be the annihilation
operators.  But then the ordering for the constraints is fixed. Almost all
terms must be such that the annihilation operator acts first, as otherwise
there is no chance for the vacuum to be a physical state. We are however free
to reorder finitely many term in the sums. Whenever we do that, we pick up some
constant from \eref{s-k-k-com}, so we can summarize all these extra terms into
a single additive {\em normal order constant}. The final result is
\beq[s-k=0]
  \op\fcl_+(0) &=& \frac 1{8\pi} \, \op P_\mu \, \op P^\mu +
                 \sumlll \op a_\mu(l) \, \op a^\mu(-l) + \hbar\, N, \zl
  \op\fcl_-(0) &=& \frac 1{8\pi} \, \op P_\mu \, \op P^\mu +
                 \sumlll \op b_\mu(-l) \, \op b^\mu(l) + \hbar\, N.
\eeq
In principle, $N$ can be different for the two constraints, but we will shortly
see that we are forced to take the same for both. Now we can hope to solve the
constraints. But still we are faced with another problem. We don't know yet
whether the classical constraint algebra \eref{s-vira} is preserved by the
quantized constraints. If we compute the commutator algebra of the constraints,
this is what we get:
\beq[s-q-vira]
  \comm{\op\fcl_+(k)}{\op\fcl_+(l)} &=& - \hbar\, (k-l) \,\op\fcl_+(k+l) -
         \hbar^2 \delta_{k+l}  \big(\ft1{12}\,D\,(k^3-k) - 2\, k \,N \big),\zl
  \comm{\op\fcl_-(k)}{\op\fcl_-(l)} &=&  \hbar\, (k-l) \,\op\fcl_-(k+l) +
         \hbar^2 \delta_{k+l}  \big(\ft1{12}\,D\,(k^3-k) - 2\, k \,N \big).
\eeq
Hence, the algebra doesn't close any more, and the freedom to choose $N$ is not
sufficient to render it closed. This means that we cannot impose the conditions
$\op \fcl_a\kt\Psi=0$ to get the physical state space. This time, we are forced
to use the ``mixed scheme'', and it is indeed possible, as we can arrange the
constraints into two complex conjugate sets, each of which forms a closed
algebra. Again we are not free to choose these sets, if we want the correct
behavior of the annihilation and creation operators. What we want is that a
state with the property
\beq[s-vac-p]
   \op a_\mu(k) \ket0 = 0 \txt{for} k<0,  \qquad
   \op b_\mu(k) \ket0 = 0 \txt{for} k>0
\eeq
becomes physical, as this would be the vacuum. By looking at the sums in
\eref{s-const}, we find that such a state satisfies $\op\fcl_+(k)\kt0=0$ for
$k<0$ and $\op\fcl_-(k)\kt0=0$ for $k>0$, because in all terms in the sums
there will be at least one $\op a_\mu$ with a negative argument, or one $\op
b_\mu$ with a positive argument, respectively, so each term will vanish
separately when acting on $\kt0$. Hence, we define the physical state condition
to be
\beq[s-q-phys]
  \op \fcl_+(k) \ket \Psi = 0  \txt{for} k\le 0, \qquad
  \op \fcl_-(k) \ket \Psi = 0  \txt{for} k\ge 0.
\eeq
The remaining constraints are the complex conjugate ones, except for $k=0$,
where the constraints are real. You can see from \eref{s-q-vira} that the two
subsets each form a closed algebra, even if we include the $k=0$ ones into both
subsets.

It is not too difficult to find a solution to these equations in form of a wave
functional. What we have to do is to solve \eref{s-vac-p}, which implies that
\eref{s-q-phys} holds for $k\ne0$, and then solve the $k=0$ constraints. Using
the definition of the $a$ and $b$ operators, \eref{s-vac-p} can be rewritten as
\beq
   \op p_\mu(k) \ket0 = \i \, |k| \, \op q(k) \ket0
            \txt{for} k \neq 0.
\eeq
The general solution is
\beq
   \Psi_0[\q] = \Phi(Q) \,
   \exp\Big( -\frac 1{4\pi\hbar} \sum_k |k| \, q_\mu(k) \, q^\mu(-k)  \Big),
\eeq
with some wave function $\Phi(Q)$ that depends on the center of mass variable
$Q_\mu=q_\mu(0)/2\pi$. When acting with the $\op\fcl_\pm(0)$ on this
functional, all terms in the sums vanish, and we are left with the Klein Gordon
equation
\beq
   (\op P_\mu \, \op P^\mu + 8 \pi \hbar \, N \big) \, \Phi(Q) = 0.
\eeq
Here we need that the two normal order constants are equal, as otherwise we
would get a contradiction.

The state given can be interpreted as the ground state of the string's internal
degrees of freedom. In this ground state the string behaves like a point
particle decribed by the wave function $\Phi(Q)$, which is completely analogous
to the relativistic point particle in the last section. However, its mass
depends on the choice we have made for the normal order constant, so
effectively it depends on the operator ordering in the constraints. The
explicit appearance of Planck's constant in $m^2=8\pi\hbar N$ means that this
is a real quantum effect. With the scalar product we have for the point
particle, we can make an ansatz for the scalar product on the string state
space. The norm of the ground states above are simply defined to be those of
the corresponding state of the point particle. If we than act on the state with
the creation operators $\op a_\mu(k)$ and $\op b_\mu(-k)$, with $k>0$, to build
up a Fock space, the scalar product can be computed by exploiting the
commutation relations.

Will this product be positive definite? Remember that we had a problem with the
signature of $g_{\mu\nu}$ in the commutator relations \eref{s-k-k-com}. For
example, we find that the norm of $v^\mu \op a_\mu(k)\kt{\Psi}$ is $\pi\hbar
k\,v^\mu v_\mu$ times the norm of $\kt\Psi$, which becomes negative if the
vector $v^\mu$ is timelike. However, here is one crucial difference to the
electo-magnetic field. If $\kt\Psi$ was a physical state, the state we just
created is not physical, as $\op a_\mu(k)$ is not an observable. To create
physical states, we have to act on the vacuum states with observables, and to
find them is of course a non-trivial task. To quantize them is even more
difficult, and it is this which leads to the famous consistency conditions of
string theory, so for example to the critical dimension, which is $D=26$ for
this model. It also fixes $N$ to be some {\em negative\/} number, which means
that the ground state of the string is a point particle with negative mass
squared. But note that this only affects the {\em observable algebra}. The
physical state space is perfectly well defined outside the critical dimension,
and everything is again manifestly gauge invariant, as we never had to impose
gauge conditions. Especially, we did not need any ghosts or whatsoever to
quantize the string. However, as already mentioned, the real problems show up
when one tries to reproduce the classical observable algebra.

\section*{Literature}
The probably nicest and most compact book on the Dirac programme is the
following collection of lectures by Dirac himself:
\begin{itemize}
\item P.A.M. Dirac, {\em Lectures on Quantum Mechanics}, Academic Press, New
York, 1965.
\end{itemize}
A more detailed review, with lots of examples but a slightly different
quantization scheme (using gauge fixing), is
\begin{itemize}
\item A.J. Hanson, T. Regge and C. Teitelboim, {\em Constrained Hamiltonian
Systems}, Accademia Nazionale dei Lincei, Roma, 1976.
\end{itemize}
At the classical level, the Hamiltonian method can be generalized in various
ways, and the most interesting one is probably the {\em De~Donder Weyl\/}
canonical formalism. Unfortunately there is, up to now, no quantum theory based
on it. You can find an introduction in
\begin{itemize}
\item H.A. Kastrup, {\em Canonical theories of Lagrangian dynamical systems in
physics}, {\em Physics Reports}, {\bf 101} (1983), 3.
\end{itemize}
All what has been done here can be reformulated in a more mathematically
rigorous way using the notions of differential geometry. This is called
``Geometric Quantization'', and here are some books on it:
\begin{itemize}
\item N.E. Hurt, {\em Geometric Quantization in Action}, Reidel, Dordrecht,
1982;
\item N.M.J. Woodhouse, {\em Geometric Quantization}, Clarendon Press, Oxford,
1992;
\item J. Sniatycki, {\em Geometric Quantization and Quantum Mechanics},
Springer Verlag, Heidelberg, 1980.
\end{itemize}
An active area where Dirac's and related methods are applied in the moment is
quantum gravity. A detailed review of the successful quantization of three
dimensional pure gravity, with emphasis on the Dirac Programme, is given in
\begin{itemize}
\item H.-J. Matschull, {\em Three dimensional canonical quantum gravity},
Topical Review, {\em Classical and Quantum Gravity}, {\bf 12} (1995), 2621.
\end{itemize}
An nice article about the problems of quantizing four dimensional gravity is
\begin{itemize}
\item C. Isham, {\em Conceptual and geometrical problems in quantum gravity},
in: {\em Recent Aspects of Quantum Fields, Proceedings, Schladming}, Springer,
Heidelberg, 1991;
\end{itemize}

\end{document}